# The Significant Role of Hydrogen in the Formation of Silicon Carbide in Evolved Stars

Guillermo Tajuelo-Castilla[1], Gonzalo Santoro[1,2,*], Lidia Martínez[1], Pablo Merino[1], José Ignacio Martínez[1], Pedro L. de Andres[1], Gary J. Ellis[3], Álvaro Mayoral[4], Ramón J. Peláez[5], Isabel Tanarro[5], Marcelino Agúndez[6], Sandra Wiersma[7], Hassan Sabbah[7], José Cernicharo[6,*], Christine Joblin[7,*], José Ángel Martín-Gago[1,*]

*[1]Instituto de Ciencia de Materiales de Madrid (ICMM), CSIC. c/ Sor Juana Inés de la Cruz 3, 28049 Madrid (Spain).*

*[2]Macromolecular Physics Department, Instituto de Estructura de la Materia (IEM), CSIC. c/Serrano 121, 28006 Madrid (Spain).*

*[3]Instituto de Ciencia y Tecnología de Polímeros (ICTP), CSIC. c/ Juan de la Cierva 3, 28006, Madrid (Spain).*

*[4]Instituto de Nanociencia y Materiales de Aragon (INMA), CSIC-UNIZAR, c/ Pedro Cerbuna 12, 50009, Zaragoza (Spain).*

*[5]Molecular Physics Department, Instituto de Estructura de la Materia (IEM), CSIC. c/Serrano 123, 28006 Madrid (Spain).*

*[6]Instituto de Física Fundamental (IFF), CSIC, c/ Serrano 113-115, 28006 Madrid (Spain).*

*[7]IRAP, Université de Toulouse, CNRS, CNES. 9 Av. du Colonel Roche, F-31028 Toulouse Cedex 4 (France).*

**Corresponding authors:**

Gonzalo Santoro: gonzalo.santoro@csic.es

José Cernicharo: jose.cernicharo@csic.es

Christine Joblin: christine.joblin@irap.omp.eu;

José Ángel Martín-Gago: gago@icmm.csic.es

**Abstract:** Cosmic dust is mainly formed in the atmospheres of evolved stars. In carbon-rich stars, amorphous carbon along with silicon carbide are the main constituents of dust grains yet the mechanisms involved in the formation of these grains are still poorly understood. Several molecular precursors have been proposed to form silicon carbide grains. Here, we have simulated in the laboratory the formation of silicon carbide dust starting from atomic C, atomic Si and $H_2$ and we have clearly identified $SiC_2$ as a key molecular precursor of nanodust analogues. We show that the interaction of molecular hydrogen with atomic carbon initiates the formation of hydrocarbons, which then react with atomic silicon to produce gas-phase $SiC_2$. In our experiments, the silicon carbide nanodust analogues are partially hydrogenated. Chemical routes for the formation of $SiC_2$ and organosilicon species are discussed on the basis of thermochemical calculations and chemical kinetics modelling. Our findings reveal the central role of molecular hydrogen in the formation of $SiC_2$ and contribute to a deeper understanding of silicon carbide dust formation processes in evolved stars, from atoms to molecules, clusters, and ultimately dust grains.

**Introduction**

Evolved stars are molecular factories and the primary source of cosmic dust, which forms in the Circumstellar Envelopes (CSEs) from molecular aggregation. In carbon-rich (C-rich) stars, dust composition consists of carbonaceous material and silicon carbide. The presence of silicon carbide grains in C-rich Asymptotic Giant Branch (AGB) stars was confirmed fifty years ago[1,2] and has been found towards a large number stars[3-5] but its formation mechanism remains unclear. The silicon chemistry in the envelope of the prototype C-rich star IRC+10216 shows SiS, SiO, $SiC_2$ and $Si_2C$ as the most abundant silicon-bearing species[3,6-10]. Although SiC has been also observed in the CSE of C-rich AGBs[3,11] it is formed in the outer layers (far from the dust nucleation region) through $SiC_2$ photodissociation and therefore does not actively contribute to silicon carbide dust formation. However, astronomical observations of $SiC_2$ towards 25 C-rich stars, have concluded that the denser the envelope, the lower the abundance of $SiC_2$ what evidences the efficient incorporation of $SiC_2$ onto dust grains. In addition, $Si_2C_2$, whose abundance is similar to that of $SiC_2$ and is restricted to regions close to the star[6], probably contributes to the formation of silicon carbide grains as well. These two species are likely the principal gas-phase molecular precursors of silicon carbide dust.

Here, we investigate the interaction of atomic carbon, atomic silicon, and molecular hydrogen through laboratory simulations in the *Stardust* machine[12,13]. Our experiments reveal the formation of silicon carbide nanodust analogues and identify $SiC_2$ as a major gas-phase species, detected only in an $H_2$-enriched atmosphere with C, Si, and $H_2$ ratios similar to those in inner envelopes of C-rich stars. Chemical modelling confirms that the reactions of Si with $C_2H$ and $C_2H_2$ drive $SiC_2$ formation. This is also the main path to $SiC_2$ in stars[6]. The analysis of the solid analogues reveals three distinct sets of nanodust analogues, namely, hydrogenated amorphous carbon (mainly aliphatic), partially crystalline hydrogenated silicon and partially hydrogenated amorphous silicon carbide. Moreover, the molecular composition analysis shows the presence of species such as SiC, $SiCH_n$, $SiC_2H_n$, $Si_2CH_n$ and $Si_2C_2H_n$, showing the very rich silicon chemistry.

## Results

### Comparison of experimental conditions with those of carbon-rich evolved stars

Although the physical conditions and chemical timescales in evolved stars are difficult to reproduce in the laboratory, our experiments emulate the relative abundances of C, Si and $H_2$ of the innermost circumstellar layers.

We used the Sputtering Gas Aggregation Source (SGAS) available at the *Stardust* machine[12,13], which vaporizes atoms from solid targets by sputtering processes. Two sputtering sources were employed: one with a C target and another with polycrystalline Si target. The atomic density of C and Si is estimated as of $2.5 \times 10^{10}$ atoms/cm$^3$ each; thus, the C/Si abundance ratio is approximately 1. This ratio corresponds to that predicted for C-rich AGBs at distances of 1-2 stellar radii[14].

We used two $H_2$ densities, which we denominate as low $H_2$ density and high $H_2$ density for consistency with our previous work[12]. In the former case, molecular $H_2$, atomic C, and atomic Si are in equal proportion, which is achieved by using the residual hydrogen in the system. The density of $H_2$ in this case is estimated at an upper limit of about $1.5 \times 10^{10}$ molecules/ cm$^3$. For the high $H_2$ density, we introduced additional $H_2$ to a density of $1.5 \times 10^{12}$ molecules/cm$^3$. In this case, the relative concentration of $H_2$ to C/Si is closer to that in CSEs (ratio ~1000) and more appropriate to simulate the chemistry in the CSEs of C-rich stars. We note that, at the experimental conditions, the pressure in the SGAS is 0.003 bar. Pressures in the inner regions of carbon-rich evolved stars can reach values up to ~ 0.001 bar at about 1 stellar radii[14].

### Structure and composition of the laboratory nanograins

Atomic Force Microscopy (AFM) images of the dust analogues shows that it consists of nanoparticles (NPs) with sizes ranging from 1 to 15 nm (Supplementary Figure 2). We note that geometrical parameters in *Stardust* condition the maximum NP size, leading to dust grains smaller than those usually found in space (typically 100 nm). The smaller grains of our experiments represent the initial stages of dust growth.

The combination of Scanning Transmission Electron Microscopy (STEM) and Electron Energy Loss Spectroscopy (EELS) provides concurrent information on the elemental composition and distribution with high spatial resolution. For the high $H_2$ density, three different types of NPs were identified: amorphous carbon, silicon and silicon carbide. Two amorphous silicon carbide NPs are shown in Figure 1, whereas carbon and silicon NPs are shown in see Supplementary

Figure 5. Figure 1a shows a TEM image of representative silicon carbide nanograins, and Figure 1b-d show false-colour elemental maps derived from EELS spectra (Supplementary Figure 3). The silicon carbide NPs exhibit carbon enriched shells of about 1 nm in thickness (Fig. 1d), attributed to the known Si depletion process, which increases superficial carbon[15].

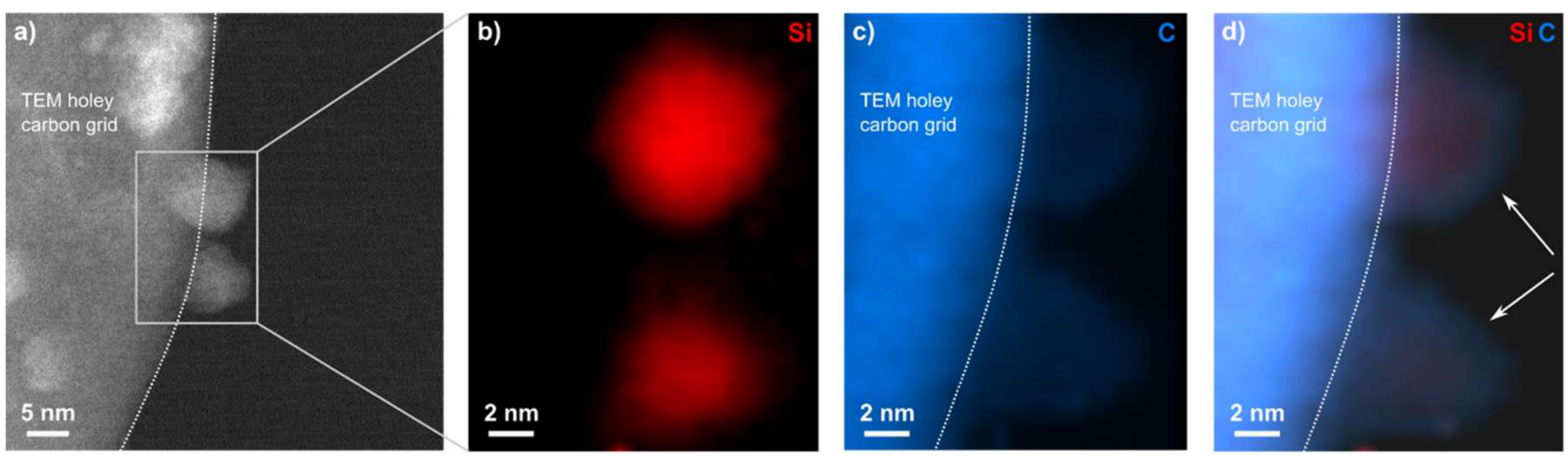

**Figure 1. Morphology and elemental distribution of silicon carbide nanodust**. a) Representative Cs-corrected STEM-HAADF image of the nanograins produced by the simultaneous vaporization of C and Si at the high H2 density conditions. b) and c) STEM-EELS chemical mapping obtained from the Si-K edge and C-K edge, respectively. d) Composite image of the elemental distribution by combining b) and c). In d) the white arrows indicate the C enriched shell of the dust analogues. The carbon holey grid used to collect the NPs for TEM measurements is indicated.

We analyzed the chemical composition of the nanodust analogues by X-ray Photoelectron Spectroscopy (XPS). Figure 2a-b presents the Si 2p and C 1s core level spectra, respectively, along with the fitting into components. The components have been assigned using our previous works and the extensive literature on epitaxial graphene on SiC with hydrogen[15,16]. The first observation is that the Si-C bond, either as silicon carbide or hydrogenated silicon carbide, became predominant upon introduction of $H_2$ (high $H_2$ density) corroborating the TEM observation of the co-existence of three NP types, i.e, amorphous carbon (a:C NPs), silicon (Si NPs) and hydrogenated silicon carbide (H:SiC NPs).

For the low $H_2$ density, the analogues are mainly composed of pure C-C/C-H and pure Si-Si/Si-H chemical bonds, indicating that, at these conditions, there is no strong interaction among carbon and silicon. Quantitative analysis (Supplementary Table 1) of the fitted spectra reveals that about 65 % of the total carbon is locked into C-C/C-H bonds and about 75 % of silicon into Si-Si/Si-H bonds. We also identified silicon carbide and hydrogenated silicon carbide as minor components.

However, when we introduce extra hydrogen into the system the relative amount of C-C/C-H and of Si-Si/Si-H in the analogues is considerably reduced and in this case the majority of carbon and silicon is locked into Si-C bonds (about 60 % of the total carbon and about 70 % of the total silicon), either neat or hydrogenated. A detailed discussion on the quantitative analysis of the XPS spectra is provided in Supplementary Section 3.

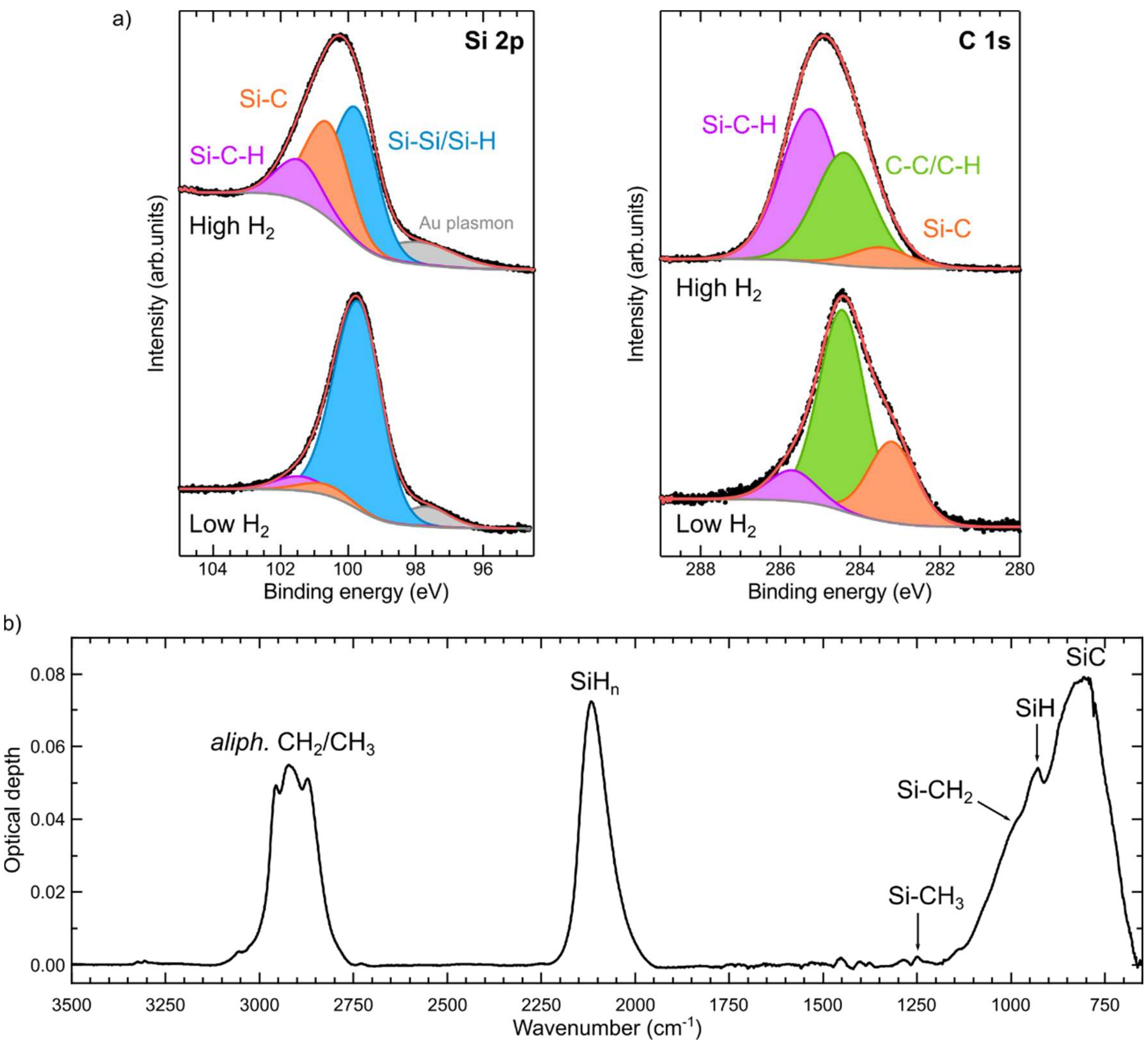

**Figure 2**. **Composition of the nanodust.** a) High resolution XPS spectra of the Si 2p and C 1s core levels of the laboratory dust nanograins. Fitting of the spectra into components is also shown. For clarity the Si $2p_{3/2}$ and Si $2p_{1/2}$ peaks are shown together although the fitting has been performed with separated contributions. The plasmon of the Au substrate is also shown. b) Baseline corrected IRRAS spectrum of the nanodust analogues synthesized at the high $H_2$ density conditions. The most important absorption features are indicated in the figure.

We also used IR spectroscopy to investigate the composition of the high $H_2$ density analogues (Fig. 2b) (see Supplementary Table 2 for band assignment). We identified silicon carbide as well as the presence of alkyl substituents into SiC (Si-$CH_2$ and Si-$CH_3$ moieties). This corroborates the XPS results that also demonstrate that a significant part of the silicon carbide bonds in the analogues are hydrogenated. In addition, we observe absorption bands associated to $SiH_n$ species and to aliphatic hydrocarbons.

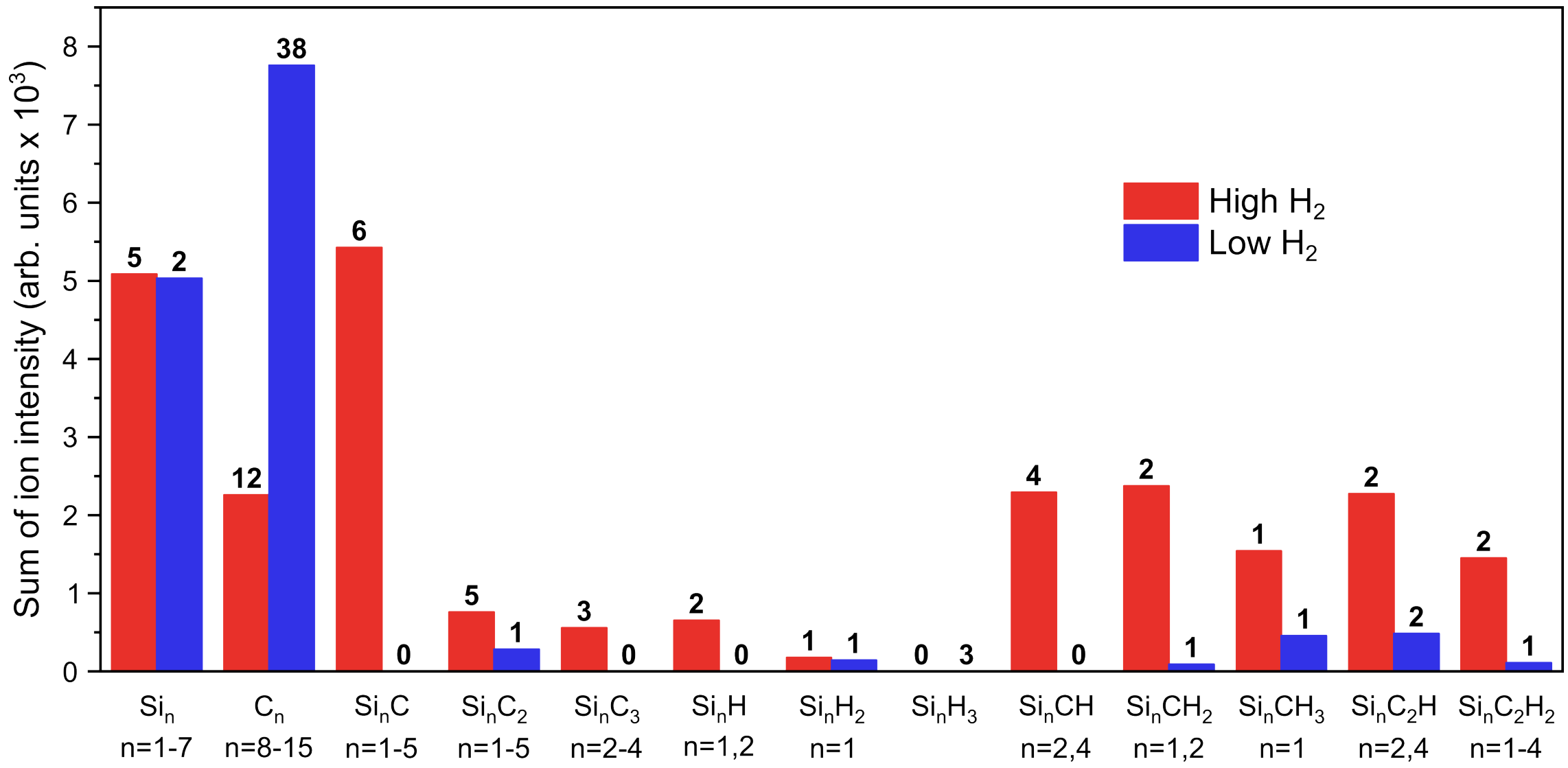

**Figure 3**. **Molecular content of the dust analogues.** Census of $Si_n$, $C_n$ and organosilicon $Si_nC_mH_l$ species (n range is indicated in the figure for each species; m=0-3; l=0-2) detected in the nanodust analogues (averaged over 2 and 4 samples for the low- and high-$H_2$ cases, respectively; see Supplementary Table 3). Bar heights correspond to the ion signal intensities measured with AROMA, and the numbers above the bars indicate the number of distinct species detected. Detailed breakdowns for $SiC_mH_l$ and $Si_2C_mH_l$ are shown in Supplementary Figure 10. Full data are available in the AROMA database (https://aroma.irap.omp.eu).

The molecular composition of the analogues was also analysed *ex situ* using Laser Desorption Ionization Mass Spectrometry (LDI-MS) at the AROMA setup[17]. We note, that the material analysed by LDI-MS (as in the XPS and IRRAS results) comes from the three types of NPs and the signal is not specific to either kind. Figure 3 shows that in the case of low $H_2$, the dominant molecular species detected are pure carbon clusters and pure silicon clusters. Upon $H_2$ addition, a number of mixed (hydro)carbon-silicon species are observed. Supplementary Figure 10 shows that $SiCH_n^+$ (n=2, 3, 5), $SiC_2H_3^+$, $Si_2CH_2^+$, $Si_2C_2H^+$ and $Si_2C_2H_2^+$ provide a fairly strong ion signal. In addition, a plethora of $SiC_nH_m^+$ species (n>1; m>0) are detected. This reflects the interaction of atomic Si with small hydrocarbons. Those are formed by chemical reactions of atomic C with $H_2$ that leads to the formation of acetylene ($C_2H_2$) and methane ($CH_4$) as stable species[12] (see next section). Previously, we have demonstrated, that the chemistry of atomic carbon and $C_2H_2$ leads to the formation of Polycyclic Aromatic Hydrocarbons (PAHs)[18] although high amounts of $C_2H_2$ with respect to $H_2$ are needed. We have also observed that this effect is enhanced in the presence of Si. These results are part of an additional study (S. Wiersma *et al.*, in preparation). The detection of species such as $SiC_9H_7^+$, likely reflects interactions between Si and PAH species such as indene or indenyl.

Overall, AFM, STEM, XPS and IRRAS spectra reveal that the nanograins are mainly composed of three types of NPs: aliphatic amorphous carbon, partially crystalline hydrogenated silicon, and amorphous silicon carbide (both hydrogenated and non-hydrogenated). This reflects the competition between different chemical pathways, as confirmed by LDI-MS which show that at low $H_2$ density, the dominant species are pure carbon and silicon clusters, in agreement with the XPS data. In contrast, high $H_2$ densities (i.e., C/Si/$H_2$ abundance ratios closer to those encountered in C-rich evolved stars) lead to the formation of a wide variety of organo-silicon species ($Si_nC_mH_l$), along with a significant fraction of silicon carbide in both its hydrogenated and non-hydrogenated forms.

**Gas-phase synthesis of molecular precursors of silicon carbide grains**

We have investigated the gas phase molecular precursors of the nanograins by Optical Emission Spectroscopy (OES) in the aggregation zone of the SGAS (Figure 4a).

For the low $H_2$ density, we detect the emission of atomic Si at 288.2 nm ($3p^2\,{}^3P$-$3p^3\,{}^3D^o$)[19] and the $C_2$ Swan band ($d^3\Pi^g - a^3\Pi^u$) at 512-517 nm. Once $H_2$ is injected into the system, Si and $C_2$ are partially consumed and we observe the emission of SiH at 410-415 nm ($A^2\Delta$-$X^2\Pi$)[20], CH at 431.2 nm (*A*–*X* band) and atomic H at 656.3 nm ($H_\alpha$ line of the Balmer series). These results are consistent with our previous works[12,21].

More importantly, at the high $H_2$ density we observe the emission of $SiC_2$ at 498.3 nm (Merrill-Sanford bands; $\tilde{A}^1B_2$-$\tilde{X}^1A_1$ system)[22-25]. The identification of $SiC_2$ is further confirmed by an additional band at 519.9 nm[22,23] (Supplementary Figure 11). For comparison, in Fig. 4a we have included the $SiC_2$ emission spectrum of the C-star IRAS 12311-3509, showing a very good agreement with our spectra. $SiC_2$ has been proposed as one of the most likely molecular precursors of silicon carbide grains.

We have carefully searched for the optical emission of $Si_2$, SiC, $Si_2C$ and $Si_2C_2$[26-31], which have also been suggested as molecular precursors of silicon carbide dust, but we have not identified any emission associated with those. Thus, although it cannot be ruled out that these species contribute to the formation of silicon carbide grains, its contribution is minor (if any) and the main molecular precursor of silicon carbide dust is $SiC_2$.

To further investigate the chemistry behind the formation of the nanodust analogues, we performed mass spectrometry at the exit of the SGAS, i.e., once the gas-phase molecular species are dragged out of the aggregation zone and no further chemistry takes place. In Fig. 4b, we compare the mass spectra obtained for the low and high $H_2$ densities for the cases when i) only C atoms are vaporized, ii) only Si atoms are vaporized and iii) both C and Si are simultaneously vaporized at both low and high $H_2$ densities. For the low $H_2$ density, gaseous species other than Ar (the sputtering gas) are not detected in any of the cases. However, for the high $H_2$ density the situation changes drastically.

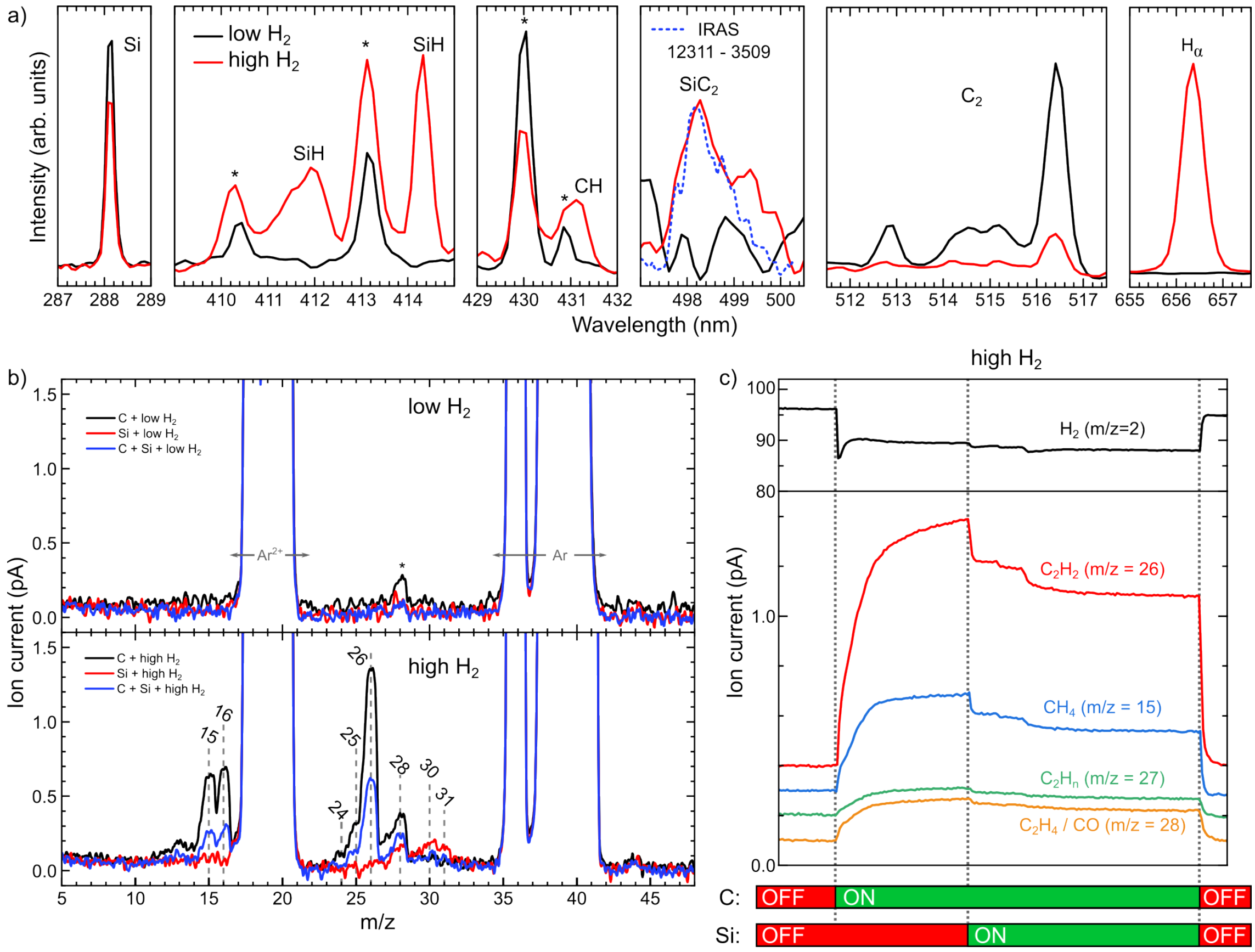

**Figure 4. Gas-phase precursors of dust analogues.** a) Optical emission spectra for low and high $H_2$ densities during the simultaneous vaporization of C and Si atoms. For comparison, the dashed line corresponds to the $SiC_2$ emission of IRAS 12311-3509 carbon star, extracted from[24]. The peaks indicated by asterisks are related to Ar, the sputtering gas. b) Mass spectra for low (upper panel) and high (lower panel) $H_2$ densities. Black lines correspond to the cases when only C atoms are vaporized, red lines correspond to the cases when only Si atoms are extracted and blue lines correspond to the simultaneous vaporization of C and Si atoms. The peak indicated by an asterisk in the upper panel is related to residual CO in the system. c) Evolution of some selected m/z values with time as C and Si sources are sequentially switched on at the high $H_2$ density conditions. The bars at the bottom indicate the status of the sources. The stepped variations after switching on the silicon source correspond to the step increases of power to avoid instabilities in the source. The curves are vertically shifted for clarity.

When only C atoms are vaporized at the high $H_2$ conditions, we detect $CH_4$, $C_2H_2$ and $C_2H_n$ at m/z = 15, 16, m/z = 26 and m/z = 27, respectively. It is quite likely that we also detect $C_2H_4$ at m/z = 28, but the residual CO in the chamber also contributes to this peak. This result is identical to that shown in our previous work[12]. On the other hand, when only Si atoms are vaporized, we detect silane ($SiH_4$) at m/z = 30, 31 as expected[21].

More interestingly, when both C and Si are vaporized simultaneously (the situation where we detect $SiC_2$ by OES), we observe that the amount of hydrocarbons is considerably diminished and $SiH_4$ is no longer detected. Thus, in this situation, additional gas-phase chemical pathways are operating that compete with those leading to aliphatics and silanes, in good agreement with the composition of the dust analogues (see previous section).

Figure 4c, presents similar results in a more compact way as only the signal at selected m/z values is depicted as function of time by sequentially switching on the sputtering sources. We note that these experiments have been designed to analyse the species that are formed/consumed by the addition of atomic Si (or atomic C, Supplementary Figure 12) to the C+$H_2$ (or Si+$H_2$, Supplementary Figure 12) mixture in the SGAS but the dust analogues presented here have been always prepared with both atomic C and atomic Si in the presence of $H_2$.

These particular experiments proceed as follows. First, a high density of $H_2$ is injected. As soon as C atoms are introduced (C sputter source is switched on), $H_2$ is partially consumed in chemical reactions leading to small aliphatic hydrocarbons ($CH_4$, $C_2H_2$, $C_2H_n$ and $C_2H_4$). Then, when Si atoms are additionally vaporized (Si sputter source is switched on after a delayed time), there is a further consumption of $H_2$ along with a decrease in the signals associated with aliphatics. This indicates that Si is interacting with the hydrocarbons or with their molecular precursors, which are mainly CH and $C_2H$[12]. If we perform this experiment in an inverse manner (Supplementary Figure 12), i.e., first we vaporize Si and at a delayed time we vaporize C in the $H_2$ atmosphere, we observe that $SiH_4$ signal vanishes once C atoms are added. Thus, C atoms are interacting with $SiH_4$ or with SiH and $SiH_2$, which are the molecular precursors of silane[21].

The results from mass spectrometry complement those from OES and point towards the importance of hydrogenated species in the synthesis of $SiC_2$.

**Discussion**

**Chemical modelling**

We have modelled the formation of $SiC_2$ considering that, according to our experimental results, hydrogenated species are actively participating in the chemistry. We have considered a set of bimolecular and association reactions whose product is $SiC_2$ and we have determined the reaction energies by density functional theory (DFT) calculations (Supplementary Table 4). We have identified numerous exergonic reactions involving SiH, $C_2H$ and $C_2H_2$. Although we have not calculated possible energy barriers, the results indicate that at least some of those reactions will contribute to the formation of $SiC_2$.

We have also constructed a chemical kinetics model based on an elemental reaction network starting from atomic Si, atomic C and $H_2$ (Supplementary Table 6) and for the high $H_2$ density conditions. The results are shown in Figure 5. The temperature of our experiments is estimated at 500 K[21] but similar results have been obtained at 1500 K (Supplementary Figure 15). We note that the set of reactions considered is not exhaustive, as the full chemistry of C+$H_2$ and Si+$H_2$ is not entirely included in the model. However, we focus on the formation routes of $SiC_2$ as this is the main species detected in the gas-phase. The chemistry of C+$H_2$ and that of Si+$H_2$ is described in our previous publications[12,21].

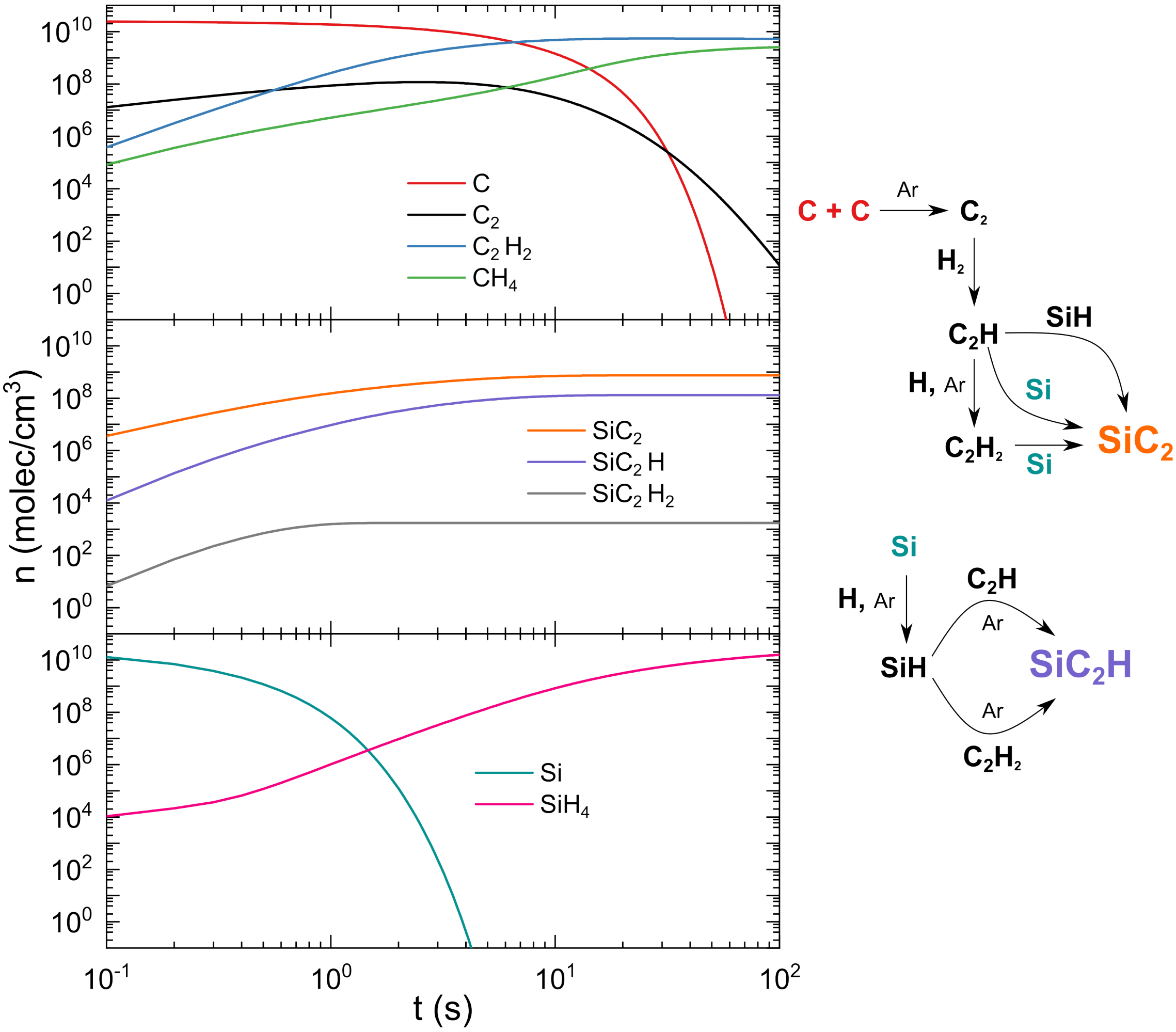

**Figure 5**. **Chemical kinetics modelling.** Temporal evolution of the abundance of selected molecules in the aggregation zone of the SGAS, according to the chemical kinetics model. The initial conditions are those corresponding to the high $H_2$ density and a temperature of 500 K. The main chemical pathways for $SiC_2$ and $SiC_2H$ are also depicted.

Our kinetics model reveals that there are three main reaction pathways. The first leads to formation of hydrocarbons and starts with the formation of $C_2$ and CH. The second involves the interaction of atomic Si and hydrogen for the formation of $SiH_n$ species. The last chemical route terminates in the formation of $SiC_2$, $SiC_2H$ and $SiC_2H_2$. These reaction pathways are in competition and once dust analogues forms from molecular aggregation they produce dust grains of hydrogenated carbon, hydrogenated silicon and silicon carbide (both neat and hydrogenated). This is also in agreement with our experimental results.

The direct formation of $SiC_2$ from Si and C atoms is mediated in our experiments by Ar, the sputtering gas, as association reactions are slow. The reaction

$$C_2H_2 + Si \rightarrow SiC_2 + H_2 \text{ (R1)}$$

despite being exoergic, has been shown by DFT calculations to present an overall energy barrier of about 0.35 eV starting from the $Si(^3P)$ ground state[32]. However, Kaiser & Gu[33] have suggested that it can proceed through intersystem crossing being fast as a bimolecular reaction even at low temperatures. Reaction (R1) has been previously considered as a possible three body reaction[32] and leading to the $SiC_2H_2$ complex if a third body is considered, as can be the case for the formation of $SiC_2$ in the photosphere of evolved stars. We have also considered this possibility (Supplementary Table 6). For the chemical kinetics modelling, we have used the experimental rates provided by Canosa et al.[34] as these are derived at buffer gas densities comparable to those of Ar in our case. From our kinetics model we identify reaction (R1) as the main route for the formation of $SiC_2$.

The following reactions also contribute notably to the formation of $SiC_2$, provided that our assumption on the rate coefficients is correct (Supplementary Table 6):

$$C_2H + Si \rightarrow SiC_2 + H \text{ (R2)}$$

$$C_2 + SiH \rightarrow SiC_2 + H \text{ (R3)}$$

$$C_2H + SiH \rightarrow SiC_2 + H_2 \text{ (R4)}$$

To our knowledge there are no data available for the rate coefficients of those reactions and we have assumed that they are of the same order of magnitude as analogous reactions involving C and CH instead of Si and SiH, respectively (see Supplementary section 9). Finally, the formation of $SiC_2H$ that, according to our model, is produced at a density similar to that of $SiC_2$, is mediated in our experiments by 3-body reactions involving Ar as only association reactions are energetically favourable (Supplementary Table 4). Kaiser & Gu[33] have established that the formation of SiCCH in the reaction of Si and $C_2H_2$ is endothermic.

**Astrophysical implications**

The results presented here could help in understanding the chemistry of silicon-carbon molecular species in AGBs and the formation of silicon carbide dust. In the inner layers of the CSE where temperature and densities permit a chemistry not far from thermodynamic equilibrium SiS, SiO, $SiC_2$, $Si_2C$, and $SiH_4$ uptake most of the available silicon. In addition to thermochemical reactions, medium/high velocity shocks that could destroy and reform species operate in the photosphere of the star.

In our experiments, we did not attempt to simulate the complete chemistry of silicon and carbon in evolved stars, but focused on Si–C(H) species as our goal was to explore plausible pathways for the formation of silicon carbide dust grains and to identify their potential molecular precursors.

In the prototypical C-rich star IRC+10216, observations of the v3=1 vibrational mode of $SiC_2$ indicate that it forms within 2-3 stellar radii[35], i.e., in the dust formation region, and we have identified it in our experiments as the main gas-phase precursor of the dust analogues. Unfortunately, no similar information at high spatial resolution is available for other stars. Medium angular resolution observations of some AGBs[36] shows a considerable increase of the $SiC_2$ abundance in the external layers of these envelopes. These observational facts, together with the result of Massalkhi et al.[3], certainly suggest that $SiC_2$ is participating in the formation of silicon carbide dust in C-rich AGB stars.

Chemical models have been developed on the basis of the reactions of Si with $C_2H_2$ and CCH[37]. In these models, $SiC_2$ is injected as a parent species into the stellar wind, although it is also efficiently formed in the external layers. From an astrophysical perspective, two distinct chemical regimes govern the formation of $SiC_2$. The first occurs in the inner layers, where thermochemical processes dominate, with temperatures ranging from a few hundred to 1500 K and densities around $10^{12}$ $cm^{-3}$. The second takes place in the outer, colder layers (T = 25 K)[38], where the lower density and kinetic temperature favour a chemistry driven by photodissociation and ionization and exothermic bimolecular reactions. These reactions lead to the formation of carbon chain radicals, cyanopolyynes, metal-bearing cations, and anions.

Silicon-carbon chains up to $SiC_6$ have been detected in IRC+10216[39] but hydrogenated species such as SiCCH, SiCCCH, $H_2CSi$, $H_2CCSi$, $H_2CSiH_2$ have not been identified yet, despite being searched for with ultrasensitive line surveys. Although they might be detected in future observations, their molecular abundances will be rather low compared with $SiC_2$ and $Si_2C$. Our LDI-MS results point to additional species that might be of interest, such as $Si_2CH_2$, $Si_2C_2H$ and $Si_2C_2H_2$.

**Conclusions**

Through laboratory astrochemistry we fabricated silicon carbide dust analogues from atomic C, atomic Si and $H_2$ with relative abundances similar to those in the CSEs of C rich AGBs. $SiC_2$ is the only gas-phase precursor identified, which is considered a key species for silicon carbide grain formation in the inner layers of the envelope of C rich stars. Our kinetic modelling shows that $SiC_2$ formation occurs through reactions Si with $C_2H_2$ and $C_2H$. Additionally, our experimental results indicate that the chemistries of C with $H_2$ and of Si with $H_2$ compete with that leading to silicon carbide formation. Our results contribute to understanding the chemistry of Si-C species in evolved stars and the formation of silicon carbide dust, for which molecular hydrogen plays an important role.

**Methods**

**Stardust configuration**

Supplementary Figure 1 shows the configuration of Stardust used during the experiments. Stardust is comprised of several different modules that can be operated in-line or off-line. Moreover, the modular concept allows for adapting the configuration of the machine to the experimental needs. A detailed description of each module is provided in[13,40,41].

In the present experiments, C atoms and Si atoms were vaporised in the MICS module and additional $H_2$ was injected through custom lateral entrances, which minimize the participation of the injected gas in the plasma discharge. Optical Emission Spectroscopy was also performed through fused silica windows in the lateral entrances of the SGAS.

Mass spectrometry was performed in the DIAGNOSIS module, which is located right after the MICS module. Once the analogues and gaseous products are dragged out of the aggregation zone, no further chemistry takes place. At DIAGNOSIS, the analogues were collected on different substrates for ex-situ AFM, TEM and LDI-MS. Due to the low deposition rate of the analogues, samples for IRRAS and XPS were also collected on suitable substrates at the DIAGNOSIS module and transferred to INFRA-ICE and ANA modules, respectively, using a UHV suitcase (base pressure $5 \times 10^{-9}$ mbar). This protocol minimized the deposition time and therefore ensured the cleanliness of the experiments.

**Synthesis of the laboratory dust.** Laboratory nanograins were fabricated using a scaled-up Multiple Ion Cluster Source MICS (form Oxford Applied Research Ltd.), a type of sputtering gas aggregation source (SGAS) equipped with 3 independent magnetrons operating in a UHV system (base pressure $1 \times 10^{-9}$ mbar). A detailed description of the MICS is provided in[41]. In the present case, we used two magnetrons loaded with a graphite target (99.95 % purity) and a polycrystalline silicon target (99.99 % purity, p doped) and ultra-pure Ar (99.999 % purity) was used as the sputtering gas with a total flow rate of 150 sccm (50 sccm through each of the 3 magnetron sources). Both loaded magnetrons were operated in Direct Current (DC) mode using typical powers of 100 W and 60 W to vaporize the graphite and silicon targets, respectively. We have estimated the atomic density from the drag current (200 mA in both cases), the kinetic energy (C:

365 eV; Si: 220 eV), and the sputtering yield[42]. Under these conditions, the density of carbon atoms and silicon atoms are similar and estimated as $2.5 \times 10^{10}$ atoms $cm^{-3}$. During the synthesis of the dust analogues, extra-pure $H_2$ (99.99% purity) was injected into the aggregation zone through the lateral entrance of the MICS by means of calibrated gas-dosing valves at flow rates of 0 sccm and 1 sccm keeping the total Ar flow rate at 150 sccm. The low and high $H_2$ density conditions correspond to 0 sccm and 1 sccm $H_2$ flow rates, respectively. We employed the following substrates for deposition of the analogues: boron-doped Si(100) with its native oxide for AFM; polycrystalline Au for IRRAS; holey carbon grids for TEM; Au(111) single crystal for XPS.

**Atomic Force Micrsocopy (AFM)**. AFM measurements were performed in the dynamic mode using a Cervantes AFM System equipped with Dulcinea electronics from Nanotec Electronica. All images were analysed using WSxM software[43]. AFM images were recorded ex situ.

**Scanning Transmission Electron Microscopy (STEM)**. TEM was carried out in a FEI Titan X-Field Emission Gun (X-FEG) operated at 300 kV, equipped with a monochromator and a CEOS spherical aberration corrector (Cs-corrector) for the electron probe assuring a spatial resolution of 0.8 Å. The microscope was operated in scanning mode (STEM) and images were collected using a High Angle Annular Dark Field detector (HAADF). The microscope was also fitted with an Oxford EDS detector and a Gatan Tridiem Energy filter for electron energy loss spectroscopy (EELS). The materials were directly deposited onto the TEM grids. TEM measurements were carried out using a cold field emission gun in a JEOL GrandARM 300 Atomic Resolution Electron Microscope that was operated at 80 kV, conditions at which we could verify that there was no electron-induce modification of the samples. The column was fitted with a JEOL double spherical Cs aberration corrector.

**X-ray photoelectron spectroscopy (XPS)**. XPS of the dust analogues was performed in situ using a Phoibos 100 1D electron/ion analyser with a one-dimensional delay line detector and a monochromatic Al Kα X-ray source (1486.6 eV). The binding energy (BE) was calibrated to the Au 4f core level peak (from the substrate) at 84 eV. As in the case of IRRAS, the deposition rate at the XPS position is low so we used the same protocol, i.e., laboratory nanograins were deposited for 2 hours at DIAGNOSIS module of *Stardust* and transferred to the XPS chamber using a UHV suitcase. For the quantification of the silicon carbide stoichiometry we have used the Si 2s and C 1s core level peaks. Both have similar sensitivity factors, and the area quantification is easier on Si 2s due to the absence of the Au plasmon (from the Au substrate) in the region of relevance. For the fitting we have used a Shirley background and the assignment of the components (for SiC, C and hydrogenated species in Si2p, C1s and Si2s core levels) have been performed using reference values from the literature and calibration samples used in previous works of the group with reference SiC crystals[15].

**Infrared Reflection Absorption Spectroscopy (IRRAS)**. IRRAS was performed in-situ in the INFRA-ICE module[41] of *Stardust* in ultra-high vacuum (UHV) conditions (base pressure at room temperature: $3 \times 10^{-10}$ mbar). Since the deposition rate at the INFRA-ICE module position is low for the sputter targets used here, laboratory dust was deposited for more than 7 hours at the DIAGNOSIS module of *Stardust* (see Supplementary Figure 1), where the deposition rate is higher, and then transferred to the INFRA-ICE module by means of a UHV suitcase ($P < 5 \times 10^{-9}$ mbar). This protocol ensures that the deposits were not air contaminated while allowing the deposition of enough material for IRRAS over a reasonable timescale. IRRAS spectra were acquired using a VERTEX 70V interferometer (Bruker) equipped with a liquid-nitrogen cooled mercury-cadmium-telluride (MCT) detector. The complete optical path is kept under vacuum ($10^{-1}$ mbar). The spectral resolution was set to 2 $cm^{-1}$ and 512 scans were co-added for each spectrum. An incidence angle of 83º was employed.

**Laser Desorption Ionization Mass Spectrometry (LDI-MS) and two-step laser desorption mass spectrometry (L2MS).** The AROMA setup[17] was employed to analyse the dust analogues using molecular analysis techniques. This setup utilizes laser desorption/ionization (LDI) methods, integrated with an ion-trap and a high-resolution time-of-flight mass spectrometer ($m/\Delta m \sim 10^4$). AROMA can investigate the molecular composition associated with solid samples through a one-step LDI or two-step laser desorption time-of-flight mass spectrometer (L2MS), the latter employing two independent lasers for desorption and ionization. L2MS significantly improves detection sensitivity for large carbonaceous molecules such as polycyclic aromatic hydrocarbons (PAHs) and fullerenes, with a detection limit as low as femtograms. The procedure involves a pulsed near-infrared laser (Nd:YAG at 1064 nm) focused on the sample to induce rapid heating for desorption, followed by a pulsed ultraviolet laser (Nd:YAG at 266 nm) that ionizes the desorbed aromatic molecules within 0.2–1.4 µs. Chemical formulas were assigned to m/z peaks with a signal-to-noise ratio over ten using the open-source mMass MS tool.

**Optical Emission spectroscopy (OES).** The light emitted by the plasma generated in front of the magnetron was collected through a fused silica window and an optical fibre of fused silica, and analysed by OES (see position in see Supplementary Figure 1) using a 193 mm focal length, motorized Czerny-Turner spectrograph (Shamrock SR-193-i-A, Andor) equipped with a CCD camera (iDus DU420A-BVF). Two diffraction gratings with 1200 grooves per mm and 1800 grooves per mm, installed in a movable turret, provide spectral ranges of 300–1200 nm and 200–950 nm, respectively, and nominal spectral resolutions of 0.22 nm and 0.15 nm, respectively (for an input slit width of 20 μm). The relative spectral efficiencies of all of the spectroscopic equipment were quantified for both diffraction gratings with a calibrated tungsten lamp.

**Mass spectrometry**. The gaseous species produced during the fabrication of the laboratory dust analogues were detected with a Quadrupole Mass Spectrometer (0–100 amu) Prisma Plus from Pfeiffer. The lowest detectable partial pressure was $10^{-10}$ mbar.

**Theory.** Gibbs free energies have been computed using Gaussian-16[44] via ab-initio Density Functional Theory, employing the UB3LYP hybrid exchange-correlation functional[45-47] and the cc-pVTZ basis set.

Kinetics have been modelled using stochastic process algebra related to a set of coupled Ordinary Differential Equations, with kinetics data for gas-phase reactions taken from the literature[43-52]. We have constructed a reaction network consisting of 67 neutral-neutral reactions to investigate plausible chemical pathways leading to the formation of $SiC_2(H)$. Supplementary Table 6 lists the reactions used in the kinetic model and the corresponding temperature-dependent rate constants. Initial conditions consisted of a gas mixture of composition $Ar/H_2/H/C/Si$, with respective densities $10^{18}/1.6\times10^{12}/10^{11}/2.5\times10^{10}/2.5\times10^{10}$ molec·cm$^{-3}$, and a temperature of 500 K. We have also evaluated the kinetics at 1500K. To assess the kinetic model, we compare the predicted time evolution of selected species with the experimental results obtained by mass spectrometry during the sequential switching on of the carbon and silicon sources.

To compare the results of the kinetic modelling under complete equilibrium conditions we have performed chemical equilibrium calculations using the same code and thermodynamic properties as those described in[53].

Briefly, the equilibrium composition is obtained by minimizing the Gibbs free energy of the mixture in the NPT ensemble,

$$G = \sum_i n_i \mu_i$$

where $n_i$ are the mole numbers and $\mu_i$ the chemical potential of species i = 1,N for an ideal gas mixture,

$$\mu_i = \mu_i^0 + k_B T \ln(p_i)$$

where $\mu_i^0$ is the standard chemical potential at temperature T, $k_B$is the Boltzmann constant, and $p_i$ is the partial pressure of the chemical species i. Standard chemical potentials are taken from[54].

The equilibrium condition is determined by searching for a minimum in the Gibbs free energy of the mixture at fixed temperature and pressure. We perform this minimization for T = 500, 1000, and 1500 K and at a constant density of $10^{18}$ molec $cm^{-3}$.

**Acknowledgements**

The authors acknowledge support from the European Research Council under the European Union's Seventh Framework Programme ERC-2013- SyG, Grant Agreement n. 610256 NANOCOSMOS (G.S., L.M., P.M, G.J.E, R.J.P., I.T., M.A., S.W., H.S., J.C., C.J., J.A.M.G.). We also thank partial funding by: Grant no. ANR-21-CE29-0001 GROWNANO (S.W., H.S., C.J.) funded by the Agence Nationale de la Recherche in France; Grants no. PID2023-146415NB-I00 (R.J.P, I.T.), PID2021-126524NB-I00 (L.M.), PID2019-106110GB-I00 (M.A., J.C.), PID2023-149077OB-C31 (J.I.M, J.A.M.G), PID2023-147545NB-I00 (M.A., J.C.), PID2023-149077OB-C33 (P.L.deA.) and PID2023-149077OB-C33 (G.J.E) funded by MICIU/AEI/10.13039/501100011033; Grant no. CNS2023-144346 (A.M.) funded by MCIN/AEI/10.13039/501100011033 and the "European Union NextGenerationEU/ PRTR"; Grants no. TEC-2024/TEC-459 SYNMOLMAT-CM (J.I.M.) funded by the Comunidad Autónoma de Madrid and co-financed by European Structural Funds; Grant No. PEJ-2021-AI/IND-21143 (G.T-C) funded by the Comunidad Autónoma de Madrid. A.M. acknowledges the use of instrumentation as well as the technical advice provided by the National Facility ELECMI ICTS node "Laboratorio de Microscopías Avanzadas" at the University of Zaragoza and to the Shanghai Key Laboratory of High-resolution Electron Microscopy at ShanghaiTech University.

The Severo Ochoa Centres of Excellence program through Grants CEX2023-001286-S (AM) and CEX2024-001445-S (G.T-C, G.S., L.M., P.M., J.I.M, P.L.deA., J.A.M-G.) is also acknowledged.

**Author contributions**

In situ experiments in *Stardust* were performed by G.T.-C., G.S., L.M and P.M.; H.S., S.W. and C.J. performed the LDI-MS experiments; L.M. performed AFM; A.M. performed STEM and EELS measurements; J.I.M. performed the DFT quantum mechanical calculations; G.J.E. participated in the interpretation of the IRRAS spectra; R.J.P and I.T. performed the OES measurements and interpretation; P.L. de A. and M.A. performed the chemical kinetics modelling. G.S. wrote the first version of the manuscript. G.S. and J.A.M.-G supervised the in situ experiments, and C.J. and J.C supervised the astrochemical interpretation. All authors discussed and contributed to the final version of the manuscript.

# Supplementary Information

## List of Contents

## Supplementary Section 1. Stardust configuration

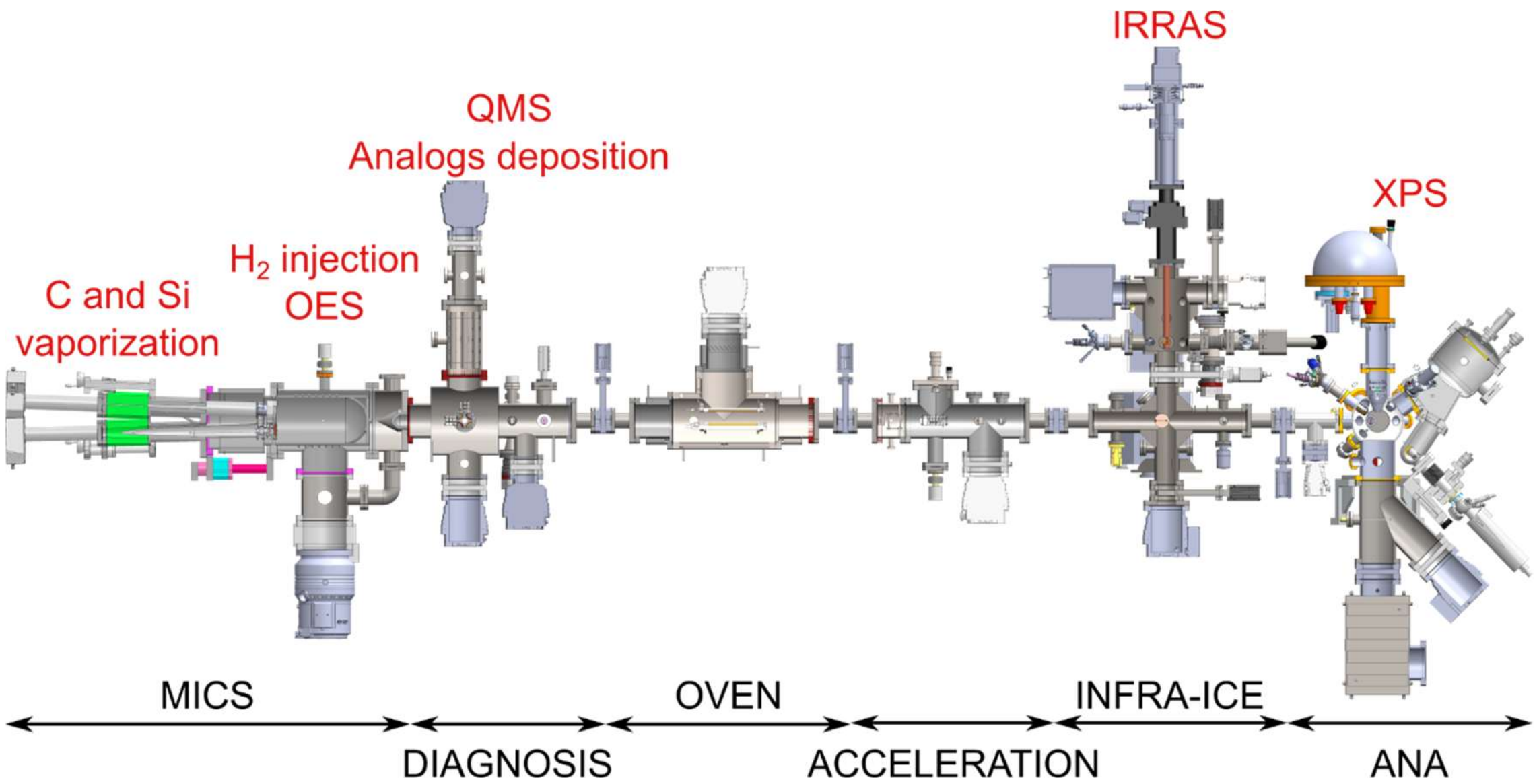

**Supplementary Figure 1**. Sketch of the STARDUST machine in the configuration used for the experiments. Reprinted and adapted from G. Santoro et al. INFRA-ICE: An ultra-high vacuum experimental station for laboratory astrochemistry. *Rev. Sci. Instrum*. **91**, 124101 (2020), with the permission of AIP Publishing.

## Supplementary Section 2. Additional characterization of the nanodust analogues by AFM, EELS and STEM

Supplementary Figure 2 presents representative atomic force microscopy (AFM) images of the nanodust analogues collected on a surface and prepared at the low and high $H_2$ density conditions. Although a full histogram representing the size distribution of the nanoparticles is difficult to obtain, we can see that there are very small particles, slightly larger than the surface roughness of the substrate (Si(111) wafer), with mean NP diameters of 2-3 nm. The features below 2 nm might be due to silicon carbide clusters. However, we note that the surface roughness of the substrate is on the range of 0.5-1 nm and therefore there is an uncertainty in the particle size statistics below 1.5 nm of diameter. These small clusters might represent the initial stages of dust analogues aggregation process. In Supplementary Section 12 we discuss the implications of our investigation on the plausible growth process of silicon carbide dust from molecules to clusters and grains.

For the high $H_2$ density conditions, the size distribution clearly moves towards higher values, indicating an enhanced aggregation mechanism promoted by the experimental abundance of $H_2$. We note that when only carbon or only silicon is vaporised at either $H_2$ conditions, narrow monomodal size distributions of NPs are obtained [1, 2].

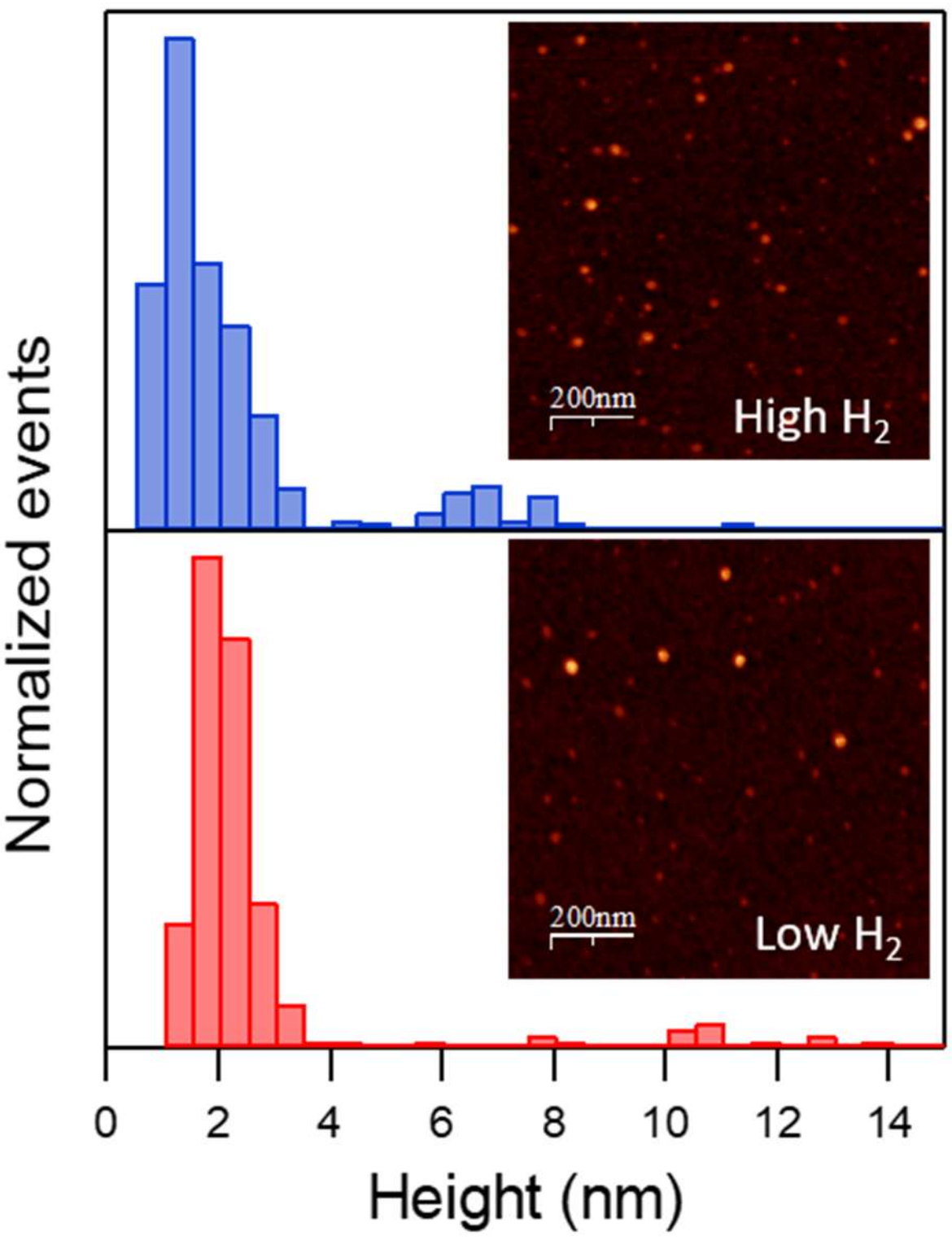

**Supplementary Figure 2**. Atomic Force Microscopy images of the nanodust analogues along with the derived histograms of nanoparticle sizes.

Supplementary Figure 3 presents the EELS spectrum of the nanoparticles of Figure 1 of the main text. The presence of oxygen in the nanodust analogues is residual. Additionally, Supplementary Figure 4 displays low magnification Cs-corrected STEM-HAADF images. The presence of dust analogues can be observed in the form of nanoparticles, as well as the rare presence of fibres (Supplementary Figure 4, left) with a lower crystallinity than the ones reported in [3].

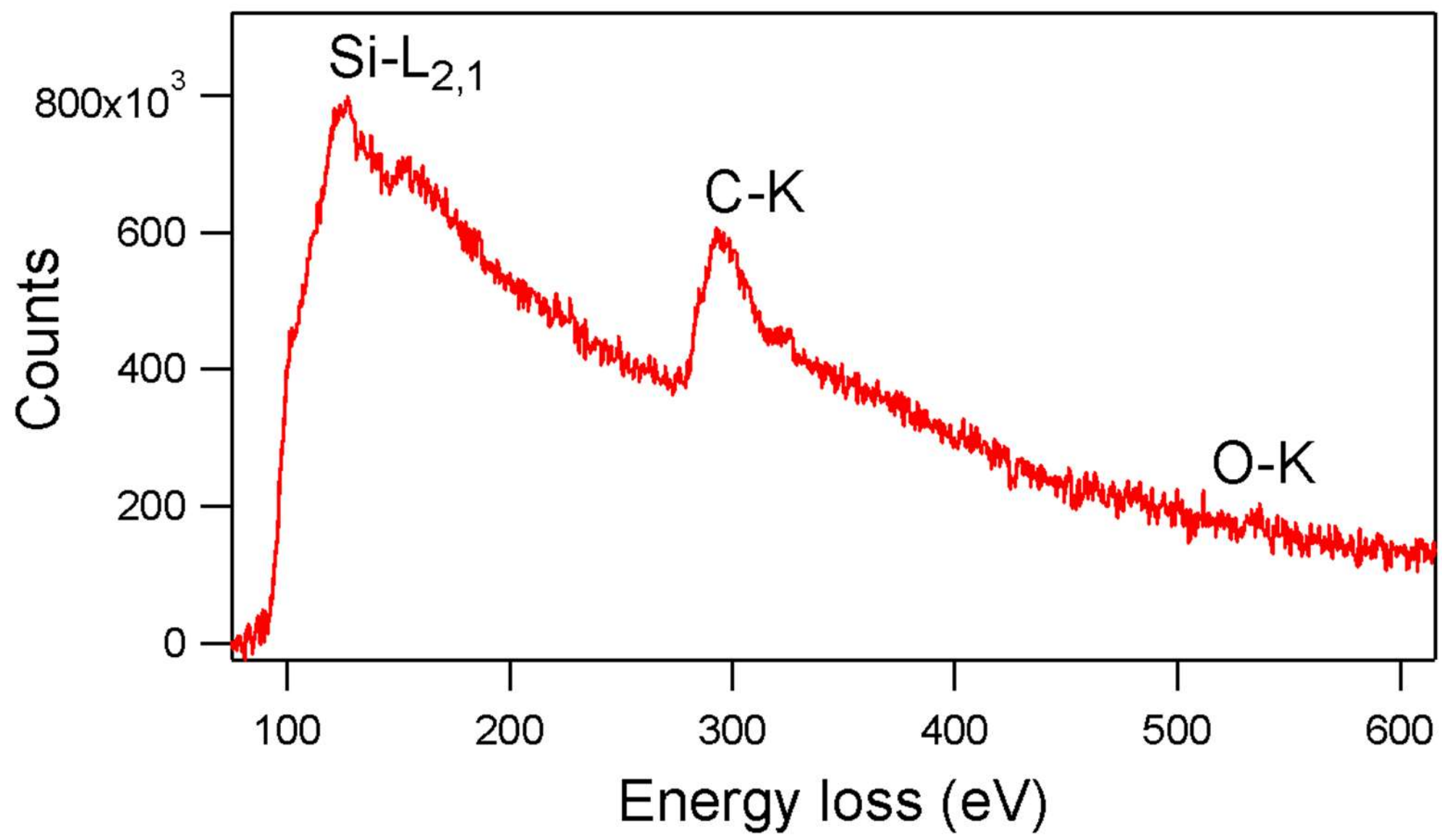

**Supplementary Figure 3.** Electron Energy Loss Spectrum (EELS) of the nanoparticles presented in Fig. 1 of the main text.

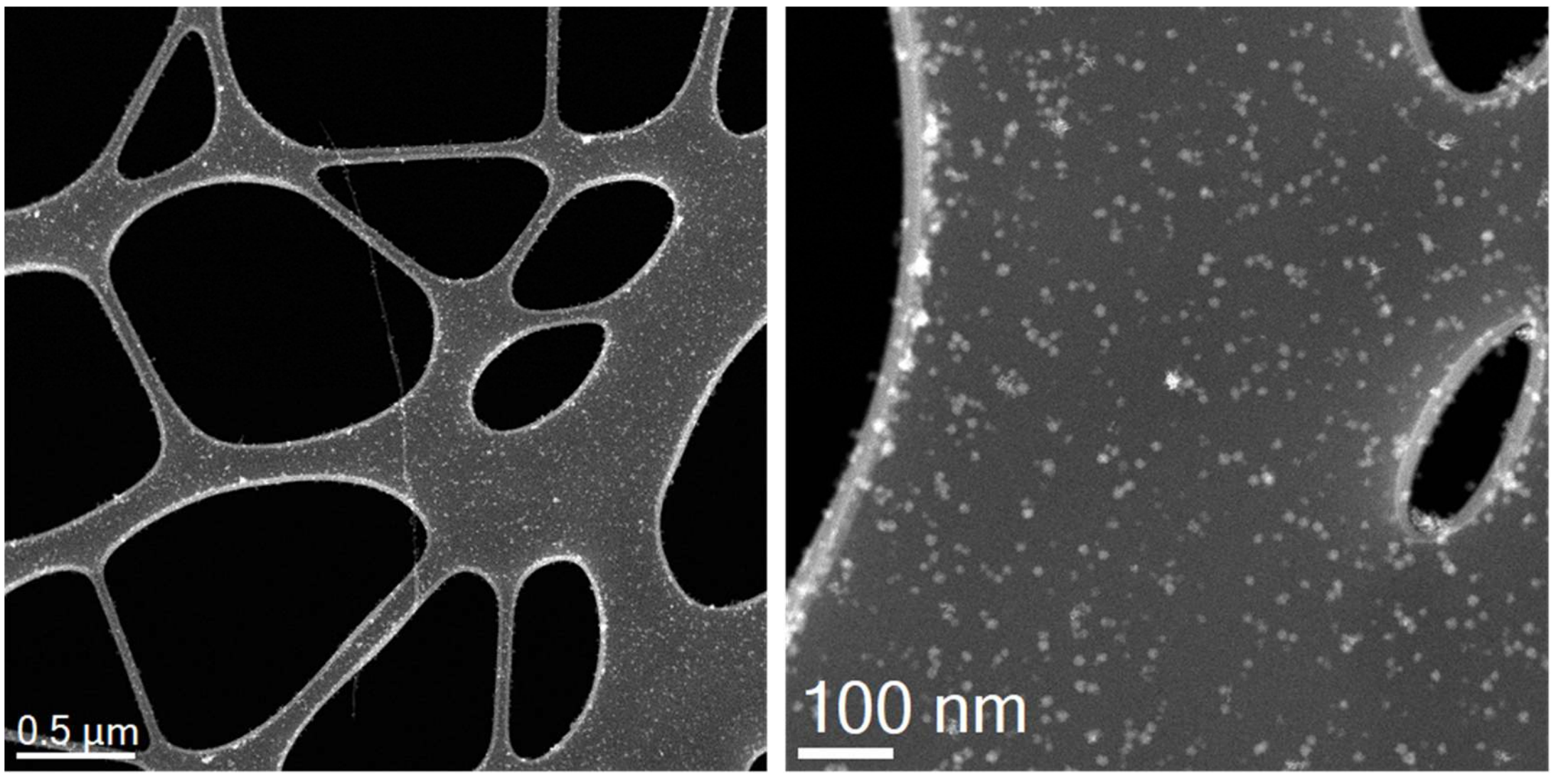

**Supplementary Figure 4.** Low magnification Cs-corrected STEM-HAADF images on a holey carbon TEM grid.

Supplementary Figure 5 presents two representative STEM images of selected nanoparticles. In the top left image, a nanoparticle with a disordered structure can be observed (which is corroborated by the corresponding FFT). These NPs are ascribed to amorphous carbon. Additionally, some nanoparticles present partially crystalline structure (Supplementary Figure 5 bottom left). In this case, the FFT displays a diffraction pattern which agrees with the presence of Si(111) domains.

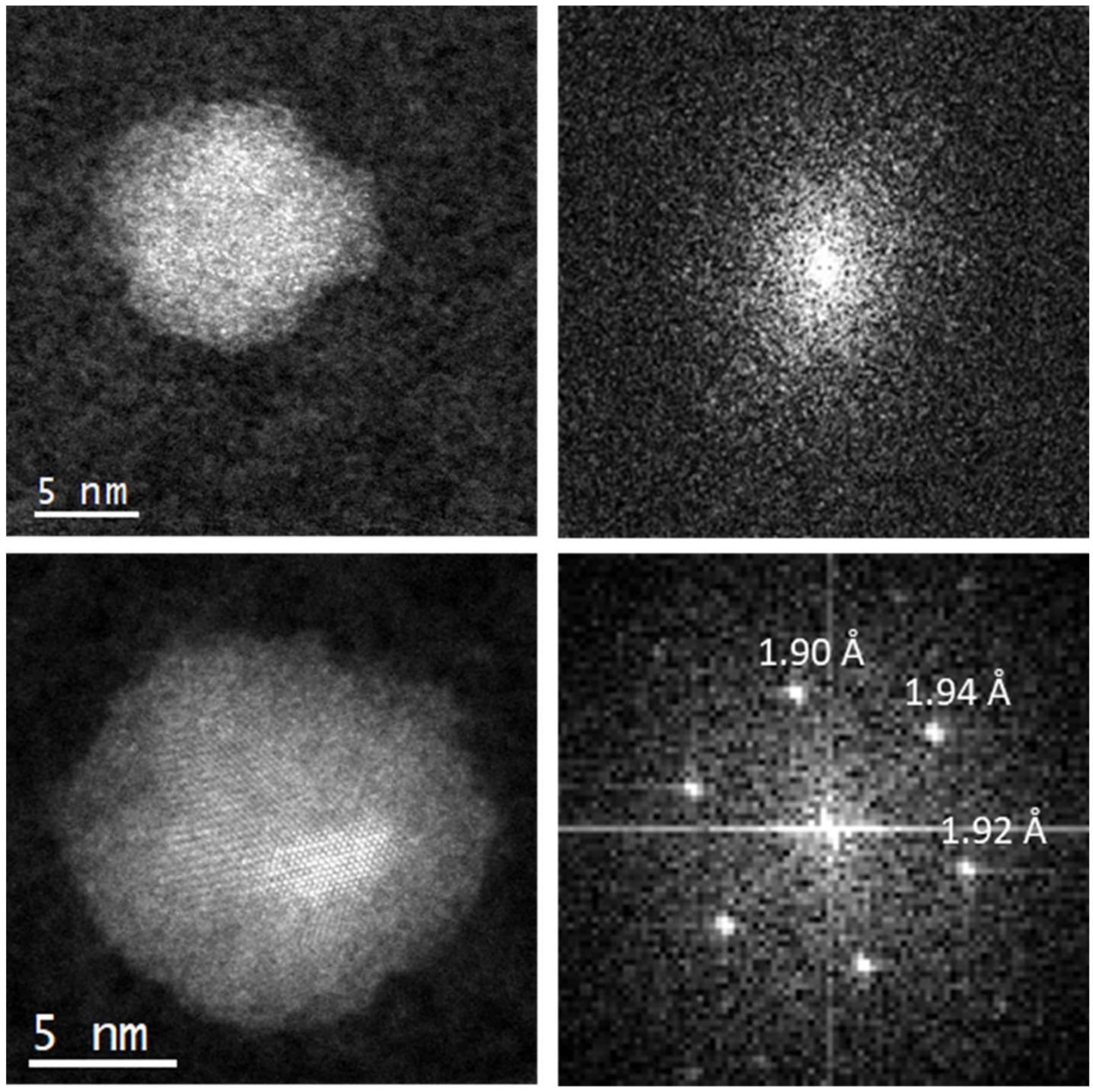

**Supplementary Figure 5.** STEM images of C nanoparticles (top left) and Si nanoparticles (bottom left) along with the corresponding FFT of the STEM images (right).

Supplementary Figure 6 presents additional images of representative laboratory nanograins along with false-colour images of the elemental distribution within the silicon carbide nanoparticles extracted from EELS spectra that complements and corroborates the results presented in Figure 1 of the main text.

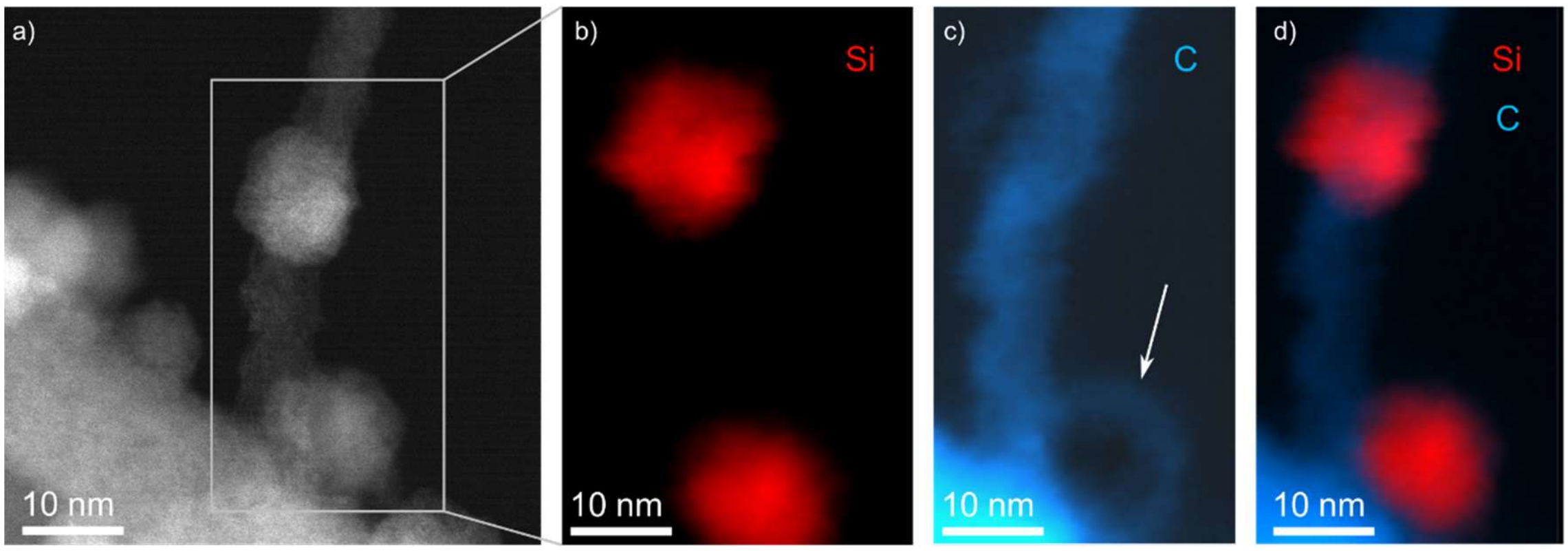

**Supplementary Figure 6**. a) Representative Cs-corrected STEM-HAADF image of the laboratory dust analogues produced by the simultaneous vaporization of C and Si at the high $H_2$ density conditions. b) and c) false-colour STEM-EELS chemical mapping obtained from the Si-K and C-K edges, respectively. d) Composite image of the elemental distribution by combining b) and c). In c) the white arrow indicates the carbon enriched shell of the nanodust analogue.

**Supplementary Section 3. Quantitative XPS results**

In this section we provide quantitative analysis of the XPS results. Supplementary Figure 7 shows survey spectra for both $H_2$ densities employed whereas Supplementary Table 1 presents a summary of the binding energies (BE) and the relative contribution of each component for the C 1s, Si 2s and Si 2p core levels.

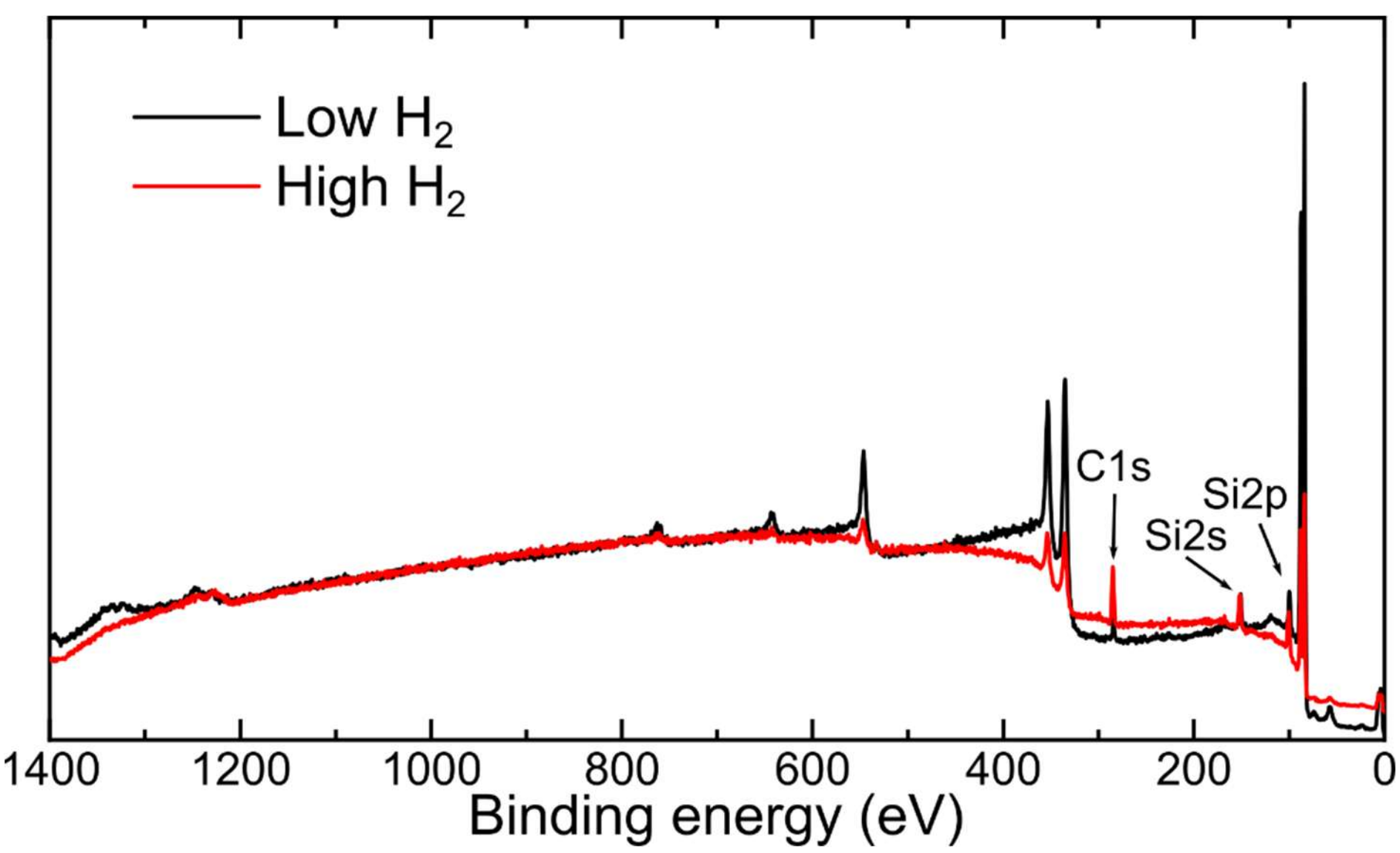

**Supplementary Figure 7**. XPS survey spectra for the low and high $H_2$ densities. C1s, Si2s and Si2p core level peaks are indicated. The rest of the peaks observed are due to the Au substrate.

**Supplementary Table 1.** Binding energy and relative contribution of each chemical environment derived from the analysis of the Si 2p and C 1s core level spectra. The binding energies (BE) provided for Si 2p corresponds to Si $2p_{3/2}$ (spin-orbit splitting: 0.6 eV). Relative contribution uncertainty of each component is estimated at 5%.

| | | Low $H_2$ density | | High $H_2$ density | |
|---|---|---|---|---|---|
| | | Peak binding energy (eV) | Relative contribution | Peak binding energy (eV) | Relative contribution |
| C1s | C-C/C-H | 284.3 | 0.63 | 284.4 | 0.43 |
| | C-Si | 283.0 | 0.25 | 283.3 | 0.10 |
| | H-C-Si | 285.4 | 0.12 | 285.4 | 0.47 |
| Si 2s | Si-Si/Si-H | 150.8 | 0.79 | 150.8 | 0.36 |
| | C-Si | 151.6 | 0.10 | 151.6 | 0.41 |
| | H-C-Si | 152.6 | 0.11 | 152.6 | 0.23 |
| Si 2p | Si-Si/Si-H | 99.4 | 0.72 | 99.5 | 0.27 |
| | C-Si | 99.5 | 0.18 | 99.9 | 0.49 |
| | H-C-Si | 101.3 | 0.10 | 101.0 | 0.24 |

References used for peak assignment [2, 4-7].

Supplementary Figure 8 shows the Si 2s high resolution XPS spectra for both $H_2$ densities along with the peak fittings into components. As we know that there are different types of nanoparticles we have taken exclusively the components related to silicon carbide, Si-C and Si-C-H, whose relative intensity is shown in Supplementary Table 1.

For the low $H_2$ density, the laboratory analogues are mainly composed of pure Si-Si/Si-H and pure C-C/C-H chemical bonds, being the normalized signal of Si twice that of C. This suggest that the silicon carbide nanoparticles present an excess of Si (Si-rich grains). Indeed, quantitative analysis leads to a Si/C ratio of 2.1 which translates into a stoichiometry of the silicon carbide NPs about $Si_2C$.

For the high $H_2$ density the calculation is *a priori* more complex as we know by TEM, that the NPs are coated by a 1 nm thick C shell. This shell attenuates the XPS Si-C signal from the NP core. However, this attenuation affects at both elements equally (with the small difference of the electron kinetic energy), and we found a ratio of Si/C = 0.7. This value, which is closer to the 1:1 stoichiometry of crystalline silicon carbide, is a result of the amorphous nature of the NPs and the possible defects/vacancies present in those. Here we note that amorphous silicon carbide is not stoichiometric but enriched in either silicon or carbon. In addition, one has to consider the uncertainty associated with the XPS quantification. In any case, the analysis clearly indicates that upon the addition of $H_2$, the composition of silicon carbide dust analogues evolves towards a close to 1:1 stoichiometry.

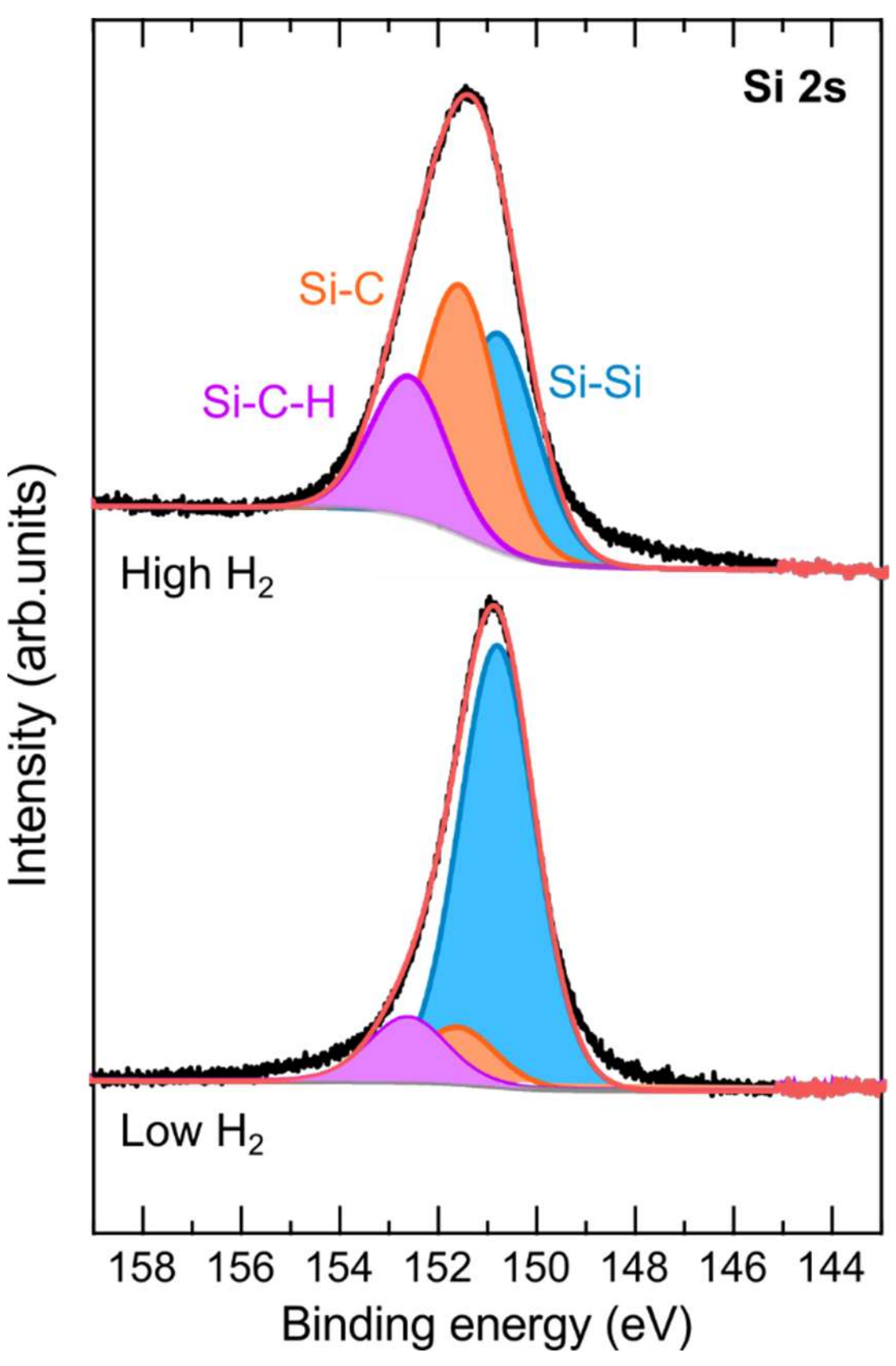

**Supplementary Figure 8**. High resolution Si 2s XPS spectra along with peak fitting into components.

## Supplementary Section 4. IRRAS additional discussion and band assignment

Due to the deposition thickness of the analogues, the as-recorded IRRAS spectrum presents interference effects that were removed by baseline correction. Supplementary Figure 9 shows both the unprocessed and baseline-corrected IRRAS spectra.

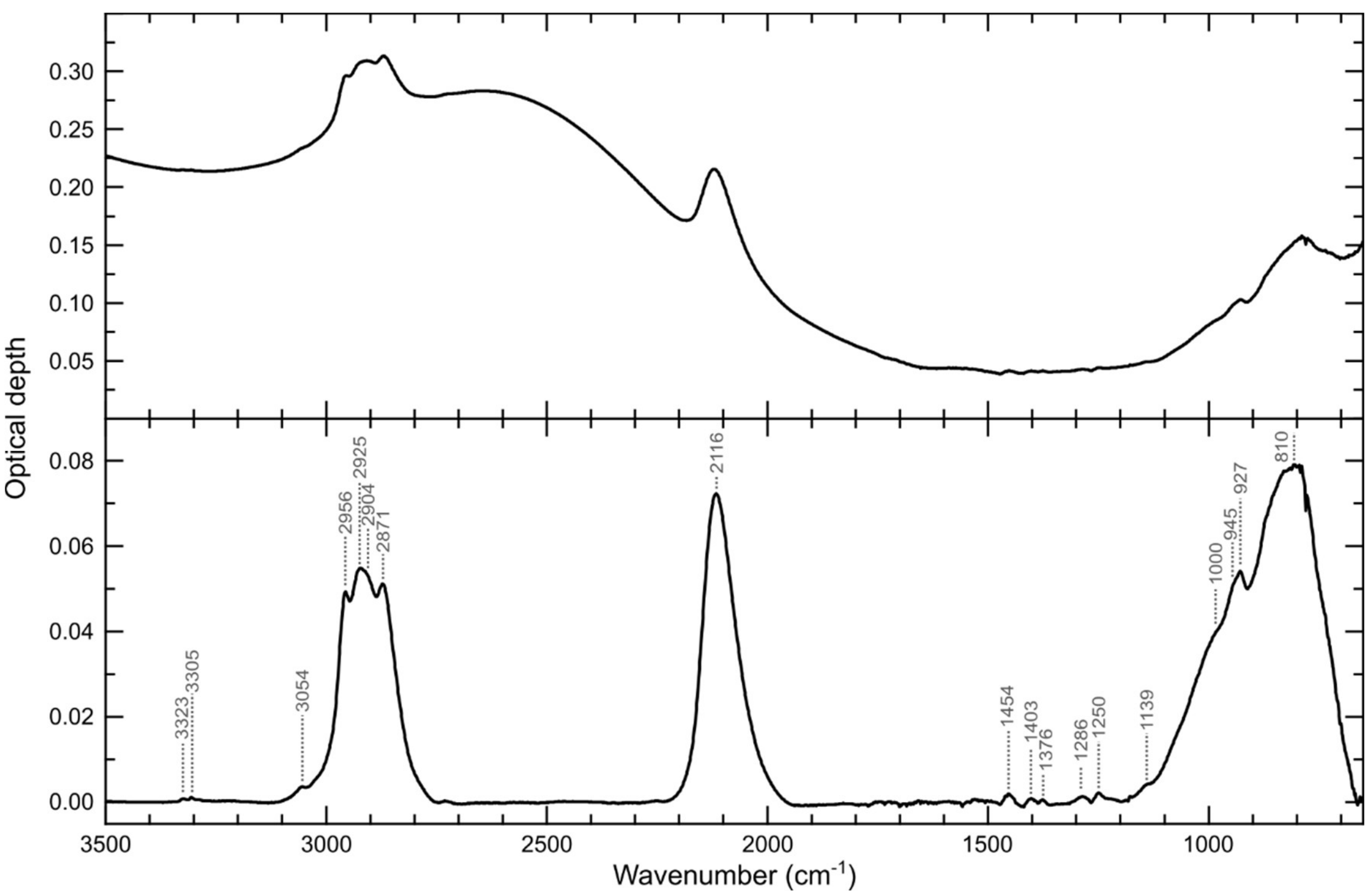

**Supplementary Figure 9**. Unprocessed (upper panel) and baseline corrected (lower panel) IR spectra of the analogues prepared at the high $H_2$ density conditions. In the lower panel, the positions of the observed IR absorption features are indicated.

Apart from the main IR features that are described in the main text, the spectrum presents some weak but interesting bands that deserve further discussion.

In particular, the two bands at 3323 $cm^{-1}$ and 3305 $cm^{-1}$ correspond to the ≡C-H stretching mode, evidencing the presence of sp-hybridised carbon in the analogues. This is in agreement with the observation of $C_2H_2$ by mass spectrometry and the plausible formation of polyynes [8]. In addition, the band at 3054 $cm^{-1}$ can be attributed to the =C-H stretching mode of $sp^2$ carbon. Although from this band alone it is difficult to infer the character of the $sp^2$ carbon, we are inclined to think that it is of olefinic nature as no indication of aromatics was observed by mass spectrometry, whereas ethylene has been most probably identified.

Despite the residual presence of sp and $sp^2$ carbon, the main carbonaceous component of the analogues consists of $sp^3$ carbon as revealed by the strong absorption features in the 3000-2800 $cm^{-1}$ region, which are due to $CH_2/CH_3$ stretching modes. Aliphatic $sp^3$ carbon is also confirmed by the small bands at 1454 $cm^{-1}$ and 1376 $cm^{-1}$, the latter corresponding to the so-called $CH_3$ "*umbrella*" mode that is very characteristic of methyl moieties.

The very strong band at 2116 $cm^{-1}$ is ascribed to SiH stretching vibrations and its asymmetric shape reveals the presence of $SiH_3$, $SiH_2$ and SiH moieties. The positions of these bands might indicate the presence of alkyl substituents on the Si atom [9] or hydrogenated amorphous silicon material [10]. SiH moieties are also responsible for the peaks at 945 $cm^{-1}$ and the shoulder at about 927 $cm^{-1}$ ($SiH_n$ bending modes).

As indicated in the main text, apart from the intense, broad feature around 810 $cm^{-1}$ that is very characteristic of the SiC stretching mode, the broad shoulder around 1000 $cm^{-1}$ is due to Si-$CH_2$ and the small features at 1250 $cm^{-1}$ and 1403 $cm^{-1}$ are characteristic of Si-$CH_3$ groups, both having the same intensity in hydrogenated amorphous silicon carbide [11]. Finally, the faint bands at 1286 $cm^{-1}$ and the shoulder at 1139 $cm^{-1}$ could be related to Si-O due to partial, low level of oxidation of the silicon component of the analogues through the residual water and oxygen in the chamber. Although this cannot be fully ruled out due to the long deposition time needed to obtain a reasonable signal in the spectrum, we are more inclined to tentatively assign these bands to Si-$CH_3$ and Si-$CH_2$-Si groups [9, 12], since the base pressure of the system is $10^{-10}$ mbar and the sample was transferred to the IR UHV chamber by means of a UHV suitcase at a pressure lower than $5 \times 10^{-9}$ mbar.

Supplementary Table 2 lists the absorption features of the analogues prepared at the high $H_2$ density conditions along with its vibrational assignment.

**Supplementary Table 2**. IR band assignment.

| Wavenumber ($cm^{-1}$) | Wavelength (μm) | Assignment [(a),(b)] | Comments |
|---|---|---|---|
| 3323 | 3.01 | $\nu$ CH (≡CH) | |
| 3305 | 3.03 | $\nu$ CH (≡CH) | |
| 3054 | 3.27 | $\nu$ CH (=CH) | Olefins/Aromatics |
| 2956 | 3.38 | $\nu_{as}$ $CH_3$ | |
| 2925 | 3.42 | $\nu_{as}$ $CH_2$ | |
| 2904 | 3.44 | $\nu_{s,F}$ $CH_2$ | |
| 2871 | 3.48 | $\nu_s$ $CH_3$ | |
| 2116 | 4.72 | $\nu$ $SiH_n$ | Asymmetric peak; convolution of SiH, $SiH_2$ and $SiH_3$ |
| 1454 | 6.88 | $\delta_{sc}$ $CH_2$, $\delta_{as}$ $CH_3$ | |
| 1403 | 7.13 | $\delta_{as}$ $CH_3$ (Si-$CH_3$) | |
| 1376 | 7.27 | $\delta_s$ $CH_3$ | |
| 1286† | 7.77 | | |
| 1250 | 8.00 | $\delta_s$ $CH_3$ (Si-$CH_3$) | |
| 1139† | 8.78 | | sh |
| ~ 1000 | 10.00 | $\gamma$ $CH_2$ (Si-$CH_2$) | sh; broad |
| 945 | 10.58 | $\delta$ $SiH_n$ | |
| 927 | 10.79 | $\delta$ $SiH_n$ | sh |
| ~ 810 | 12.35 | $\nu$ SiC | Broad band |

(a) The vibrational modes are abbreviated as follows: *ν*: stretching; *δ*: deformation (b: bend; sc: scissor); *γ*: wagging; *s*: symmetric; *as*: asymmetric; *ip*: in-plane; *oop*: out-of-plane; *sh*: shoulder.

(b) Assignments from Refs. [9-14].

†Due to the uncertain assignment of these bands, we have not ascribed them to specific molecular groups (see text).

## Supplementary Section 5. Molecular content analysis of the nanodust analogues

**Supplementary Table 3**. List of samples used for the molecular content analysis using Laser Desorption Ionization Mass Spectrometry (LDI-MS) and two-step laser desorption mass spectrometrometry (L2MS) in the AROMA setup. Deposition time corresponds to the duration of sample collection, and mass refers to the resulting mass of the deposited nanodust layer. Sample names match the datasets available in the AROMA database (http://aroma.irap.omp.eu).

| database names | $H_2$ density | time (min) | mass (μg $cm^{-2}$) |
|---|---|---|---|
| ST_Si-C-2 | Low | 80 | 1.16 |
| ST_Si-C-3 | Low | 80 | 0.92 |
| ST_Si-C-H2-3 | High | 40 | 34.8 |
| ST_Si-C-H2-4 | High | 40 | 36.9 |
| ST_Si-C-H2-5 | High | 40 | 36.9 |
| ST_Si-C-H2-2 | High | na | na |

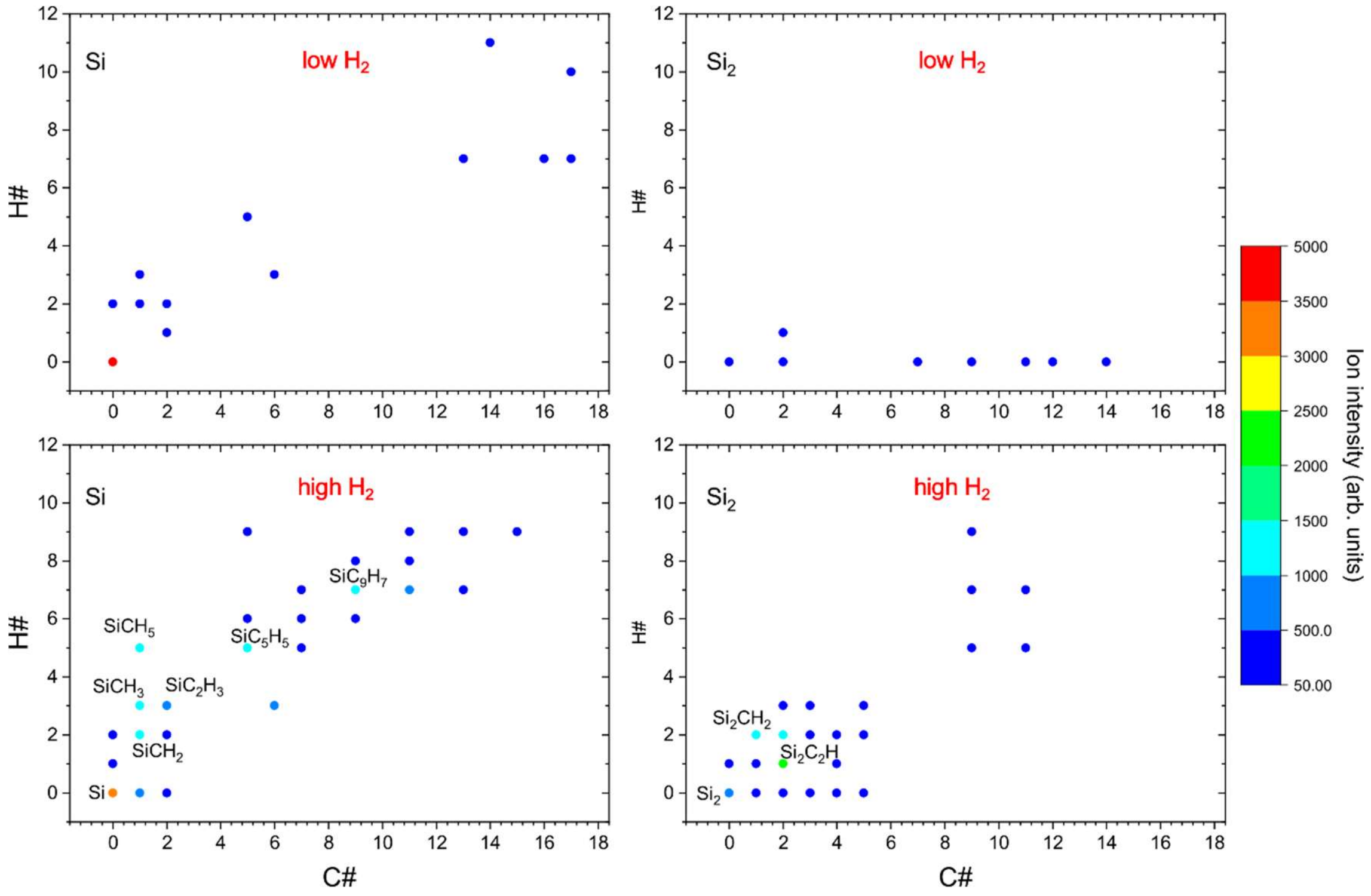

**Supplementary Figure 10.** Plot of hydrogen number (H#) versus carbon number (C#) for each identified species containing Si (left panel) or $Si_2$ (right panel) in their chemical formula. The four graphs represent samples grown with Si + C at low (upper panel) and high (lower panel) $H_2$ levels. The coloured scale indicates the peak intensities recorded in the mass spectra.

## Supplementary Section 6. Optical emission spectroscopy at different $H_2$ flow rates

Supplementary Figure 11 shows the optical emission spectra at selected spectral ranges for different flow rates of $H_2$ injected into the aggregation zone of the SGAS. The $H_2$ low density conditions corresponds to 0 sccm (no extra $H_2$ introduced into the system) whereas 1 sccm corresponds to the high $H_2$ density conditions. The relative variation of the intensity of selected peaks with the $H_2$ flow rate is also shown.

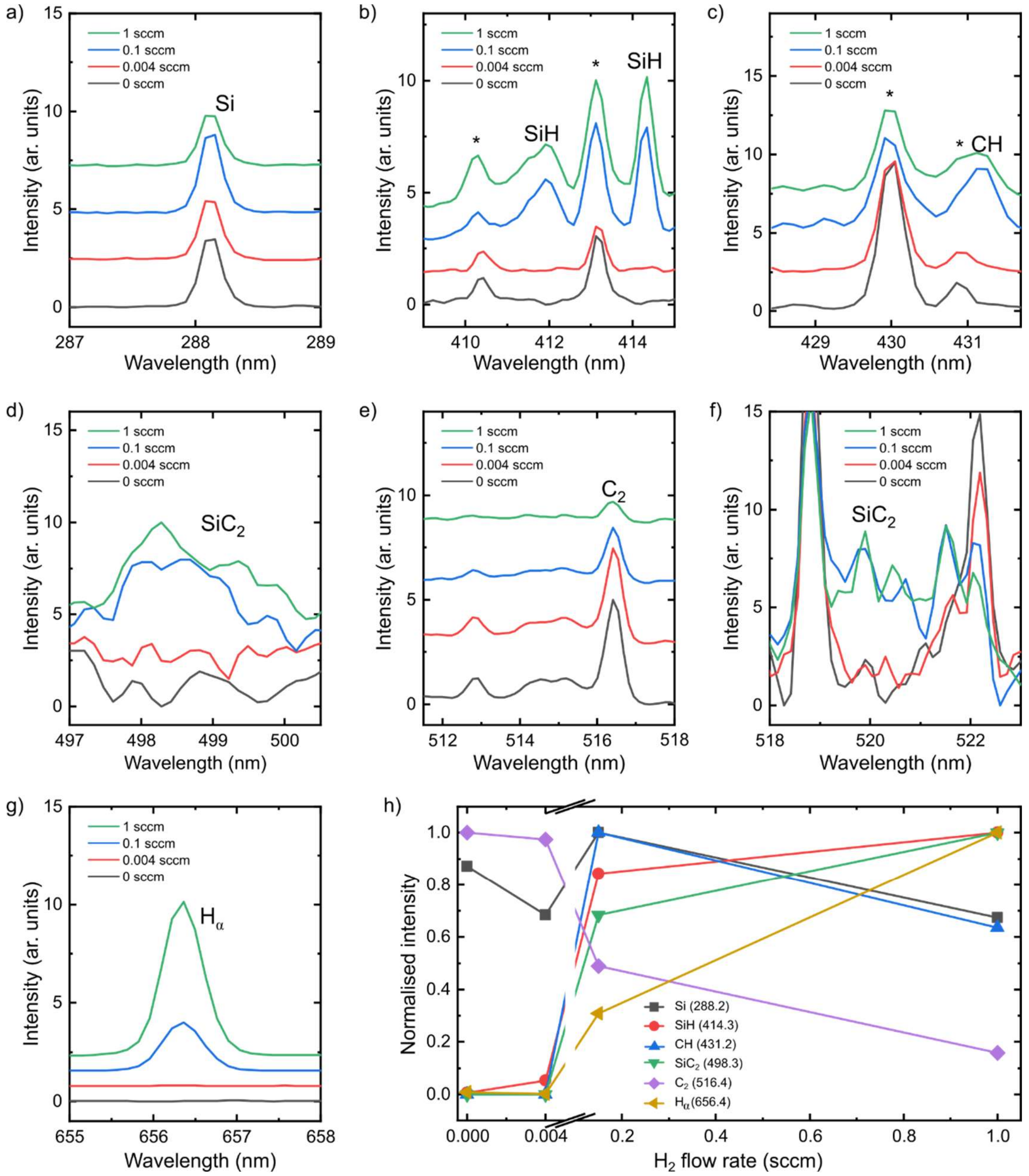

**Supplementary Figure 11**. a-g) OES spectra during the simultaneous vaporization of C and Si atoms for different $H_2$ flow rates injected into the aggregation zone of the SGAS of Stardust. The peaks indicated by asterisks are related to Ar, the sputtering gas. h) Normalized intensity of some selected OES peaks with the flow rate of $H_2$. The wavelength of the peaks used are indicated in nm in the figure.

## Supplementary Section 7. Evolution of gas-phase molecules as Si and C sources are sequentially switched on

To complement the results shown in Figure 4 of the main text, in Supplementary Figure 12 we present the evolution of some selected m/z values during the sequential switching on of the Si and C sources. First, $H_2$ is injected at the high $H_2$ density conditions, subsequently atomic Si is introduced and afterwards C atoms are vaporised. As can be observed, once C atoms are vaporised, silane ($SiH_4$) is consumed and hydrocarbons are formed.

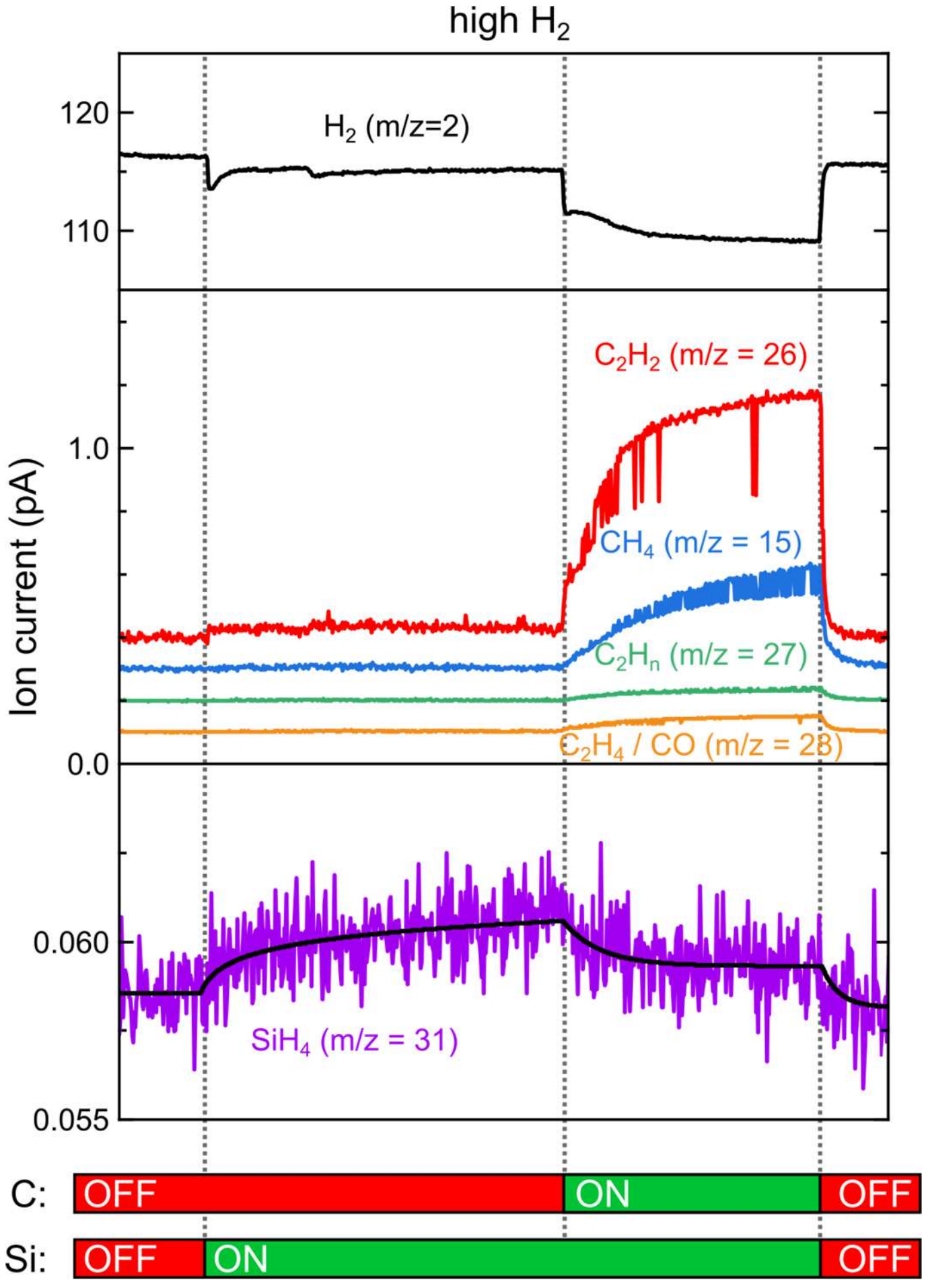

**Supplementary Figure 12**. Evolution of some selected m/z values with time as C and Si sources are sequentially switched on. The bars at the bottom indicate the status of the sources. The stepped variations after switching on the silicon source correspond to the step increases of power to avoid instabilities in the source. Spikes in the signals are ascribed to instabilities in the ionization filament of the mass spectrometer. In the lower panel the black line is a guide to the eye. The curves are vertically shifted for clarity.

**Supplementary Section 8. Calculated reaction energies for selected reactions leading to $SiC_2$**

To investigate the feasibility of $SiC_2$ formation from $SiH_n$ and $C_2H_n$ species, we carried out DFT quantum mechanical calculations (theory level: UB3LYP + cc-pVTZ). Below we list the reactions investigated along with the associated Gibbs-free energy variation (ΔG) at 500 K. Many of the selected reactions are exergonic.

**Supplementary Table 4**. Variation of the Gibbs-free energy for selected reactions leading to SiC, $SiC_2$ and hydrogenated organosilicon compounds. Exergonic reactions are highlighted.

| Reaction | ΔG (kcal/mol) |
|---|---|
| **$Si + C \rightarrow SiC$** | **-87.55** |
| **$Si + C_2 \rightarrow SiC_2$** | **-138.71** |
| $Si + C_2 \rightarrow SiC + C$ | 41.20 |
| **$Si + C_2H \rightarrow SiC_2 + H$** | **-35.25** |
| **$SiH + C_2 \rightarrow Si + C_2H$** | **-40.10** |
| **$SiH + C_2 \rightarrow SiC_2 + H$** | **-76.29** |
| **$SiH + C_2 \rightarrow SiC_2H$** | **-59.94** |
| $Si + C_2H \rightarrow SiC + CH$ | 72.49 |
| $SiH + C_2H \rightarrow SiH_2 + C_2$ | 39.59 |
| **$SiH + C_2H \rightarrow SiC_2 + H_2$** | **-66.36** |
| $SiH + C_2H \rightarrow SiC_2H + H$ | 2.48 |
| $SiH + C_2H \rightarrow SiCH_2 + C$ | 27.82 |
| $SiH + C_2H \rightarrow SiCH + CH$ | 39.89 |
| **$SiH + C_2H \rightarrow SiC_2H_2$** | **-98.72** |
| **$SiH + C_2H \rightarrow Si + C_2H_2$** | **-56.65** |
| $Si + C_2H_2 \rightarrow SiH + C_2H$ | 52.65 |
| **$Si + C_2H_2 \rightarrow SiC_2 + H_2$** | **-13.71** |
| $Si + C_2H_2 \rightarrow SiC_2H + H$ | 55.13 |
| $Si + C_2H_2 \rightarrow SiCH_2 + C$ | 80.46 |
| $Si + C_2H_2 \rightarrow SiCH + CH$ | 92.54 |
| **$Si + C_2H_2 \rightarrow SiC_2H_2$** | **-46.08** |
| $Si + C_2H_2 \rightarrow SiH_2 + C_2$ | 92.23 |

**Supplementary Section 9. Isovalent/isoelectronic chemistry of silicon and carbon**

As mentioned in the main text, we have assumed that the rate coefficients for reactions R2, R3 and R4 are on the same order of magnitude as analogous reactions involving C and CH instead of Si and SiH. Although this is commonly done [15], to support the isoelectronic/isovalent assumption we have computed the Gibbs reaction free-energies at the UB3LYP/cc-pVTZ level of theory. The results are presented in Supplementary Table 5. These data demonstrate that the silicon-bearing reactions are similarly exoergic to their carbon-substituted counterparts, lending thermochemical support to the adopted rate coefficients.

**Supplementary Table 5**. Comparison of the variation of the Gibbs-free energy of analogous reactions involving silicon and carbon

| Reaction | ΔG (kcal/mol) | Reaction | ΔG (kcal/mol) |
|---|---|---|---|
| $C_2H + Si \rightarrow SiC_2 + H$ (R2-Si) | -35.25 | $C_2H + C \rightarrow C_3 + H$ (R2-C) | -59.34 |
| $C_2 + SiH \rightarrow SiC_2 + H$ (R3-Si) | -76.29 | $C_2 + CH \rightarrow C_3 + H$ (R3-C) | -91.13 |
| $C_2H + SiH \rightarrow SiC_2 + H_2$ (R4-Si) | -66.36 | $C_2H + CH \rightarrow C_3 + H_2$ (R4-C) | -81.14 |

The gas-phase Gibbs reaction free-energies show that the formation of both $SiC_2$ and $C_3$ clusters is thermodynamically favoured for all the considered reactions, with all ΔG values being largely negative. A systematic comparison between silicon- and carbon-driven pathways shows that the carbon analogous reactions (R2-C, R3-C and R4-C) are more exergonic than the corresponding reactions involving silicon (R2-Si, R3-Si and R4-Si), reflecting the higher intrinsic thermodynamic stability of pure carbon clusters with respect to mixed Si-C species. Nevertheless, the reactions involving silicon and silicon species remain highly favourable, particularly when hydrogenated species are involved. Specifically, R3-Si and R4-Si, which proceed via SiH, are significantly more exergonic than R2-Si, which only involves atomic Si, highlighting the key stabilizing role of hydrogen in silicon-carbon chemistry. In both the Si and C reaction sets, channels leading to H elimination are more exergonic than those producing $H_2$, indicating that radical-rich environments provide an additional thermodynamic driving force for cluster growth.

Importantly, the Si reactions are expected to be highly viable from a thermokinetic perspective. They involve open-shell reactants (Si, SiH, $C_2$, $C_2H$) and lead directly to the particularly stable $SiC_2$ molecule. The strong exergonicity, the radical nature of the entrance channels, and the experimental observation that hydrogenated species are rapidly formed and consumed suggest that the activation barriers for R2-Si, R3-Si, and R4-Si reactions are low, making them competitive and efficient under hydrogen-rich gas-phase conditions.

**Supplementary Section 10. Kinetic chemical network and modelling**

Figure 4c in the main text shows results where the C source is first switched on and after a delay Si atoms are injected into the system. The experimental delay amounts to hundreds of seconds, which we set in our simulations as t=100 s as a mere qualitative indication. The main observations are: (1) Immediately after injecting C the density of $H_2$ decays by about 10%. (2) the main production is $C_2H_2$, followed by $CH_4$ in a ratio of about 3:1, (3) $C_2H_4$ and $C_2H_3$ are produced marginally at a rate of about 2:1. (4) After Si is injected, the amount of $H_2$ decays slowly by about 1 %. (5) On the other hand, $C_2H_2$ and $CH_4$ both decay by about 20 %, while there is a very small variation for $C_2H_4$ and $C_2H_3$.

By comparison, the model predicts the following (Supplementary Figure 13): (1) Immediately after injecting C the density of $H_2$ decays by about 2 %. (2) the main production is $C_2H_2$, followed by $CH_4$ in a ratio of about 6:1. (3) $C_2H_4$ and $C_2H_3$ are produced marginally in a rate of about 2:1. (4) After Si is injected, the amount of $H_2$ decays slowly by about 1%. (5) Both $C_2H_2$ and $CH_4$ decay by about 2 %, while the small variation for $C_2H_4$ and $C_2H_3$ is negligible. Finally, the model predicts the production of $SiC_2$ with an abundance comparable to that of $CH_4$. The qualitative agreement is very good showing that the set of reactions selected, although incomplete, suffices for the kinetic modelling of the process.

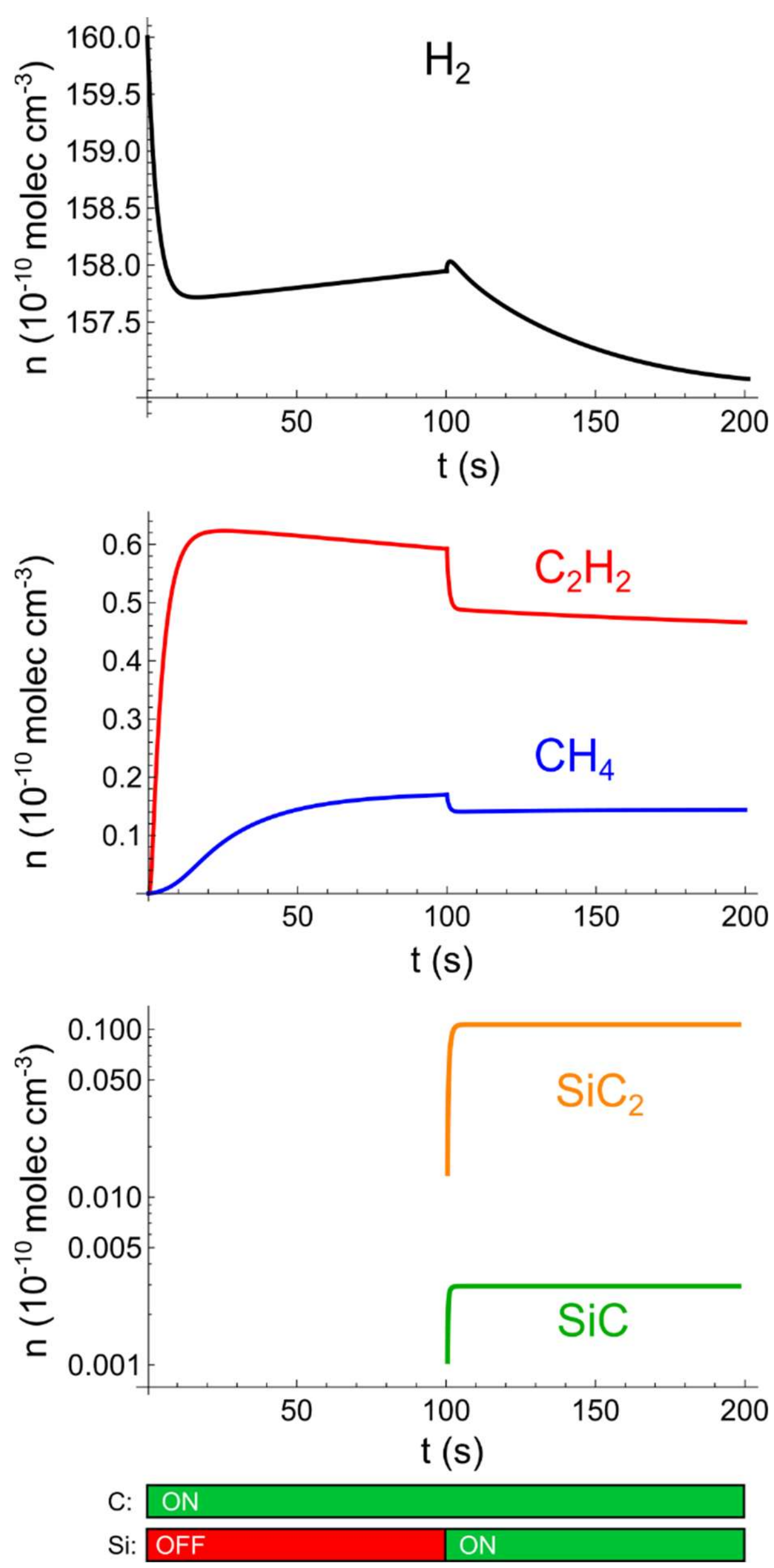

**Supplementary Figure 13.** Evolution of the density of selected species according to the kinetic model for the case when first C atoms are injected into a $H_2$ atmosphere (high $H_2$ density case) and Si atoms are injected into the system at a delayed time of 100 s. The results qualitatively reproduce the experimental results for the species detected by mass spectrometry (Fig. 4 of main text).

On a complementary set of experiments Si is injected first followed then by C vaporisation (Supplementary Figure 12).

Experimentally we observe the following (Supplementary Figure 12): (1) Immediately after injecting Si the density of $H_2$ decays by about 1 %, followed by a larger depletion of about 5 % after C is injected. (2) $SiH_4$ is formed, although in a small quantity. (3) $C_2H_2$ and $CH_4$ are formed after C is injected in a similar way to the former sequence. In comparison, the model predicts (Supplementary Figure 14): (1) a small decay of $H_2$ by about 0.5 % followed by 2 % after C is injected. (2) Production of $SiH_4$, but with a negligible variation after C is injected. (3) Production of $C_2H_2$ and $CH_4$ in a similar way to in the former sequence. (4) Finally, production of $SiC_2$ and $SiC_2H$ at comparable abundances.

The numerical accuracy in the forward propagation of the network was closely monitored by evaluating the sum rule related to the mass conservation. The simulation is stopped when the sum rule is not sufficiently well fulfilled (which always happens for larger times than those considered here).

Finally, we have also evaluated the effect of temperature on the kinetics. Supplementary Figure 15 shows the results of the kinetic modelling at 1500 K. The results are very similar to those at 500 K, the main difference being that the chemistry is faster, as expected. We want to note that that $SiC_2$ and $SiC_2H_2$ formation are favoured at high temperatures, whereas $SiH_4$ is preferentially produced at lower temperatures. The latter is related to the higher efficiency of the $SiH_4$ dissociation into $SiH_2$ and $H_2$ (see also section S11). This reaction is exothermic by 89.9 kcal/mol but requires enough thermal energy to overcome the dissociation barrier.

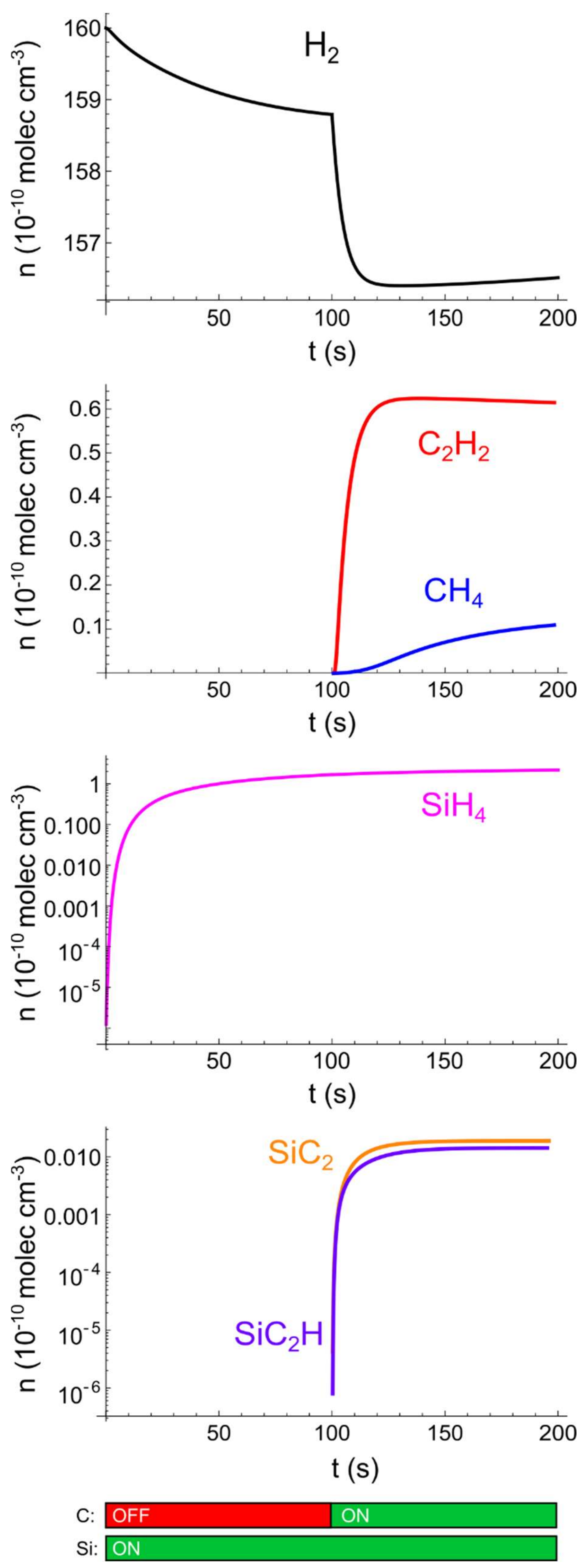

**Supplementary Figure 14.** Evolution of the density of selected species according to the kinetic model for the case when first Si atoms are injected into a $H_2$ atmosphere (high $H_2$ density case) and C atoms are injected into the system at a delayed time of 100 s. The results qualitatively reproduce the experimental results for the species detected by mass spectrometry (Fig. S12).

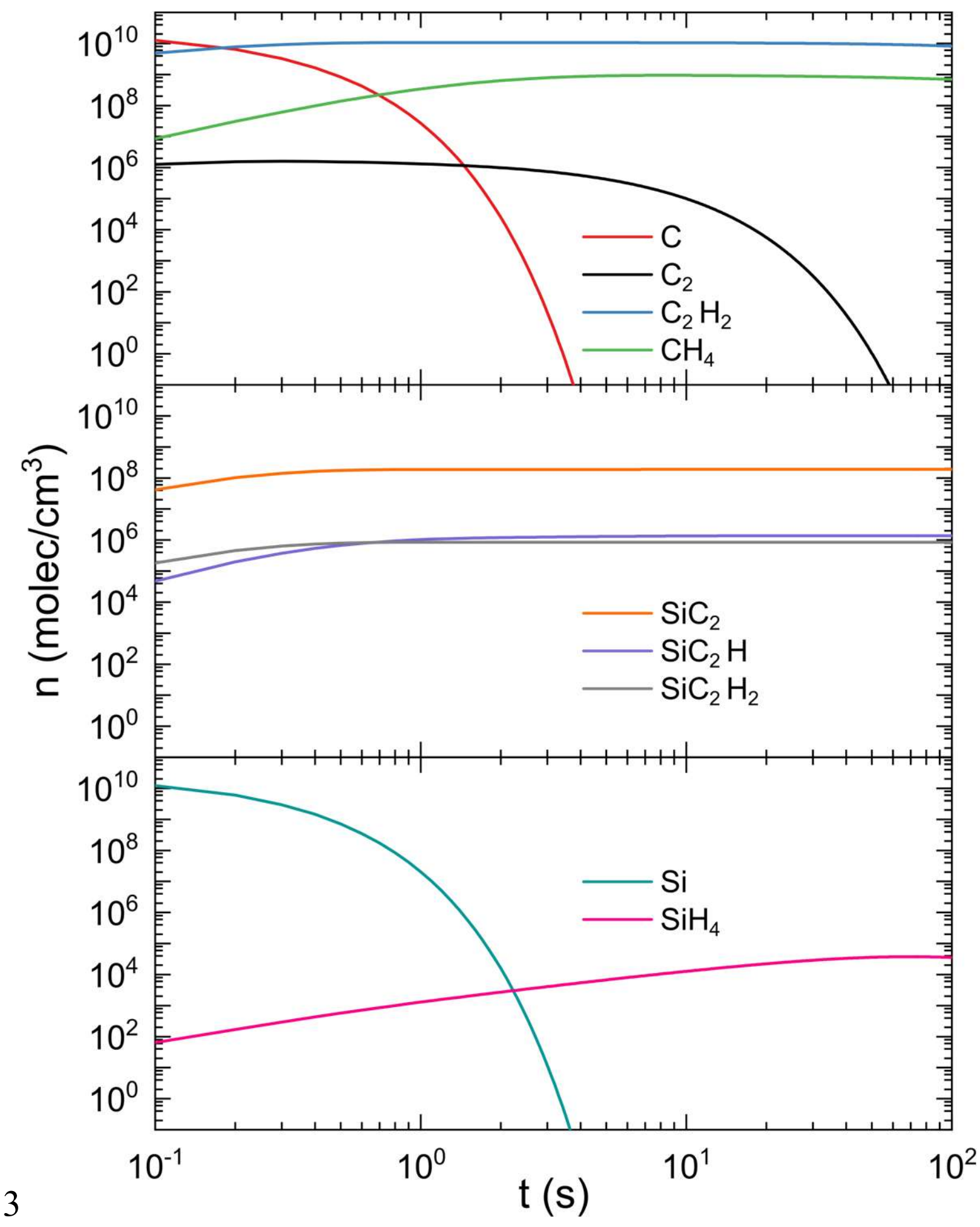

3

**Supplementary Figure 15**. Temporal evolution of the chemical composition of selected species in the aggregation zone of the SGAS, according to the chemical kinetics model developed. The initial conditions are those corresponding to the high $H_2$ density and a temperature of 1500 K.

**Supplementary Table 6**. Reactions used in the kinetic model along with temperature-dependent rate constants.

$k = A\left(\frac{T}{298\ K}\right)^n e^{-B/RT}$. A in $s^{-1}$ (o=1), $cm^3s^{-1}molecule^{-1}$ (o=2) and $cm^6s^{-1}molecule^{-1}$ (o=3). B (kJ/mol).

| i | A | n | B | R | ref |
|---|---|---|---|---|---|
| 1 | 1.e-10 | 0 | 0 | $C_2H_2 + Si \rightarrow C_2Si + H_2$ | [16] |
| 2 | 1.e-11 | 0 | 0 | $C_2H + Si \rightarrow C_2Si + H$ | [17] |
| 3 | 1.e-11 | 0 | 0 | $C_2 + HSi \rightarrow C_2Si + H$ | [17] |
| 4 | 1.e-11 | 0 | 0 | $C_2H + HSi \rightarrow C_2Si + H_2$ | [17] |
| 5 | 7.e-10 | 0 | 97 | $C + H_2 \rightarrow CH + H$ | [18] |
| 6 | 3.e-10 | 0 | 14 | $CH + H_2 \rightarrow CH_2 + H$ | [19] |
| 7 | 5.e-11 | 0 | 30 | $CH_2 + H_2 \rightarrow CH_3 + H$ | [20] |
| 8 | 2.e-11 | 0 | 57 | $CH_3 + H_2 \rightarrow CH_4 + H$ | [21] |
| 9 | 2.e-10 | 0 | 0 | $C + CH_4 \rightarrow C_2H_2 + H_2$ | [22] |
| 10 | 1.e-32 | -1 | 3 | $H + H + Ar \rightarrow H_2 + Ar$ | [23] |
| 11 | 1.e-10 | 0 | 402 | $H_2 + Ar \rightarrow H + H + Ar$ | [24] |
| 12 | 3.e-32 | -6 | 0 | $CH + H + Ar \rightarrow CH_2 + Ar$ | [25] |
| 13 | 5.e-31 | -2 | 0 | $C + C + Ar \rightarrow C_2 + Ar$ | [25] |
| 14 | 9.e-13 | 0 | 0 | $C + CH_2 \rightarrow C_2H + H$ | [26] |
| 15 | 3.e-10 | 0 | 46 | $CH_2 + CH_2 \rightarrow C_2H_2 + H + H$ | [27] |
| 16 | 2.e-10 | .3 | 0 | $C_2H + H \rightarrow C_2H_2$ | [28, 29] |
| 17 | 1.e-12 | 0 | 0 | $C_2H_2 + C \rightarrow CH_3 + H$ | [26] |
| 18 | 3.e-30 | 0 | 6 | $C_2H_2 + H \rightarrow C_2H_3 + Ar$ | [24] |
| 19 | 2.e-28 | .2 | 0 | $C_2H_3 + H + Ar \rightarrow C_2H_4 + Ar$ | [28] |
| 20 | 7.e-29 | 0 | 0 | $H + Si + Ar \rightarrow HSi + Ar$ | [30] |
| 21 | 2.e-33 | 0 | 0 | $H_2 + Si + Ar \rightarrow H_2Si + Ar$ | [30] |
| 22 | 6.e-33 | 0 | 0 | $H + HSi + Ar \rightarrow H_2Si + Ar$ | [30] |
| 23 | 5.e-28 | 0 | 0 | $H_2Si + H + Ar \rightarrow H_3Si + Ar$ | [30] |
| 24 | 2.e-30 | -3 | 7 | $H_2 + H_2Si + Ar \rightarrow H_4Si + Ar$ | [30] |
| 25 | 5.e-30 | 0 | 0 | $C + Si + Ar \rightarrow CSi + Ar$ | [26] |
| 26 | 8.e-30 | 0 | 0 | $C_2 + Si + Ar \rightarrow C_2Si + Ar$ | [26] |
| 27 | 1.e-10 | 0 | 0 | $CH_4 + Si \rightarrow CH_2Si + H_2$ | [16] |
| 28 | 1.e-14 | 0 | 0 | $C + H_4Si \rightarrow H_3Si + CH$ | [31] |

**Supplementary Table 6 (continue)**. Reactions used in the kinetic model along with temperature-dependent rate constants.

$k = A\left(\frac{T}{298\,K}\right)^n e^{-B/RT}$. A in $s^{-1}$ (o=1), $cm^3s^{-1}molecule^{-1}$ (o=2) and $cm^6s^{-1}molecule^{-1}$ (o=3). B (kJ/mol).

| i | A | n | B | Reaction | ref |
|---|---|---|---|---|---|
| 29 | 1.e-14 | 0 | 0 | $CH_4 + Si \rightarrow HSi + CH_3$ | [32] |
| 30 | 1.e-12 | 0 | 29 | $CH_3 + H_4Si \rightarrow H_3Si + CH_4$ | [33] |
| 31 | 7.e-14 | 0 | 0 | $C + H_2 \rightarrow CH_2$ | [33] |
| 32 | 2.e-12 | 0 | 0 | $CH + H_2 \rightarrow CH_3$ | [33] |
| 33 | 1.e-12 | 0 | 0 | $CH_2 + H_2 \rightarrow CH_4$ | [33] |
| 34 | 3.e-14 | 0 | 0 | $C_2 + H_2 \rightarrow C_2H + H$ | [33] |
| 35 | 5.e-13 | 0 | 163 | $C_2H_2 + H_2 \rightarrow C_2H_4$ | [29] |
| 36 | 1.e-14 | 0 | 0 | $H_2 + HSi \rightarrow H_3Si$ | [33] |
| 37 | 4.e-12 | 0 | 62 | $H_2 + H_3Si \rightarrow H_4Si + H$ | [34] |
| 38 | 7.e-10 | 0 | 0 | $CH + H \rightarrow C + H_2$ | [35] |
| 39 | 4.e-28 | 0 | 22 | $C + H + Ar \rightarrow CH + Ar$ | [35] |
| 40 | 5.e-31 | 0 | 0 | $CH_2 + H + Ar \rightarrow CH_3 + Ar$ | [35] |
| 41 | 4.e-16 | 3 | 36 | $CH_4 + H \rightarrow CH_3 + H_2$ | [36] |
| 42 | 9.e-12 | 0 | 0 | $CH_2 + CH_3 \rightarrow C_2H_4 + H$ | [20, 24] |
| 43 | 9.e-23 | -9 | 42 | $CH_3 + H + Ar \rightarrow CH_4 + Ar$ | [37] |
| 44 | 7.e-12 | 0 | 47 | $C_2H_4 + CH_3 \rightarrow C_2H_3 + CH_4$ | [24, 29] |
| 45 | 4.e-11 | 0 | 0 | $CH_2 + CH_2 \rightarrow C_2H_4$ | [20] |
| 46 | 3.e+14 | 0 | 321 | $H_2Si \rightarrow HSi + H$ | [30] (o=1) |
| 50 | 3.e+15 | 0 | 292 | $H_3Si \rightarrow H_2Si + H$ | [30] (o=1) |
| 51 | 3.e-11 | 2 | 1 | $H_2Si + H \rightarrow HSi + H_2$ | [30] |
| 52 | 5.e-32 | 0 | 0 | $C_2H_2 + Si + Ar \rightarrow C_2H_2Si + Ar$ | [38] |
| 53 | 1.e-10 | 0 | 0 | $C_2H + HSi \rightarrow C_2HSi$ | [17] |
| 54 | 8.e-11 | 0 | 13 | $H_4Si + H \rightarrow H_3Si + H_2$ | [39, 40] |
| 55 | 2.e-32 | -4 | 10 | $H_2Si + H + Ar \rightarrow H_3Si + Ar$ | [30] |
| 56 | 3.e-11 | 0 | 0 | $H_3Si + H_3Si \rightarrow H_4Si + H_2Si$ | [41] |
| 57 | 4.e-12 | 2 | 6 | $H_3Si + H \rightarrow H_2Si + H_2$ | [33] |
| 58 | 1.e-8 | 0 | 19 | $H_4Si \rightarrow H_2Si + H_2$ | [42] (o=2) |
| 59 | 1.e-13 | 3 | 50 | $H_4Si + Si \rightarrow H_2Si + H_2Si$ | [43] |

**Supplementary Table 6 (continue)**. Reactions used in the kinetic model along with temperature-dependent rate constants.

$k = A\left(\frac{T}{298\,K}\right)^{n} e^{-B/RT}$. A in $s^{-1}$ (o=1), $cm^{3}s^{-1}molecule^{-1}$ (o=2) and $cm^{6}s^{-1}molecule^{-1}$ (o=3). B (kJ/mol).

| i | A | n | B | Reaction | ref |
|---|---|---|---|---|---|
| 60 | 2.e-12 | 1 | 92 | $H_4Si + Si \rightarrow H_3Si + HSi$ | [43] |
| 61 | 3.e-10 | 0 | 2 | $CH_2 + Si \rightarrow CSi + H_2$ | [16] |
| 62 | 3.e-10 | 0 | 268 | $CH_2 \rightarrow C + H_2$ | [44] |
| 63 | 3.e-10 | 0 | 375 | $CH_2 \rightarrow CH + H$ | [45, 46] |
| 64 | 3.e-8 | 0 | 595 | $C_2 \rightarrow C + C$ | [47] |
| 65 | 1.e-10 | -0.16 | 169 | $Si + C_2 \rightarrow CSi + C$ | [48] |
| 66 | 1.e-29 | 0 | 0 | C + CSi + Ar → C2Si + Ar | [48] |
| 67 | 1.e-10 | 0 | 3 | $H_2Si + C_2H_2 \rightarrow$ Products | [49, 50] |

**Supplementary Section 11. Equilibrium distribution of species in the kinetic network**

The equilibrium condition is determined by searching for a minimum in the Gibbs free energy of the mixture at fixed temperature and pressure. We perform this minimization for T = 500, 1000, and 1500 K and at a constant density of $10^{18}$ molec cm$^{-3}$. The results are shown in Supplementary Figures 16, 17 and 18.

Although thermodynamic equilibrium is not expected to be reached in our experiments, the equilibrium calculations provide an important insight. They indicate that $SiC_2$ formation is favoured at high temperatures, whereas $SiH_4$ is preferentially produced at lower temperatures (Supplementary Figures 16, right), as the comparison of the kinetic modelling at 500 K and 1500 K also show. As already mentioned in Supplementary Section 10, these trends can be rationalized by considering the different free energies involved in those processes. For example, the exothermic reaction $SiH_4 \rightarrow SiH_2 + H_2$ (89.9 kcal/mol) is expected to deplete $SiH_4$ at high temperatures once the thermal energy is sufficient to overcome the dissociation barrier, explaining the decline of $SiH_4$ abundance as the temperature increases.

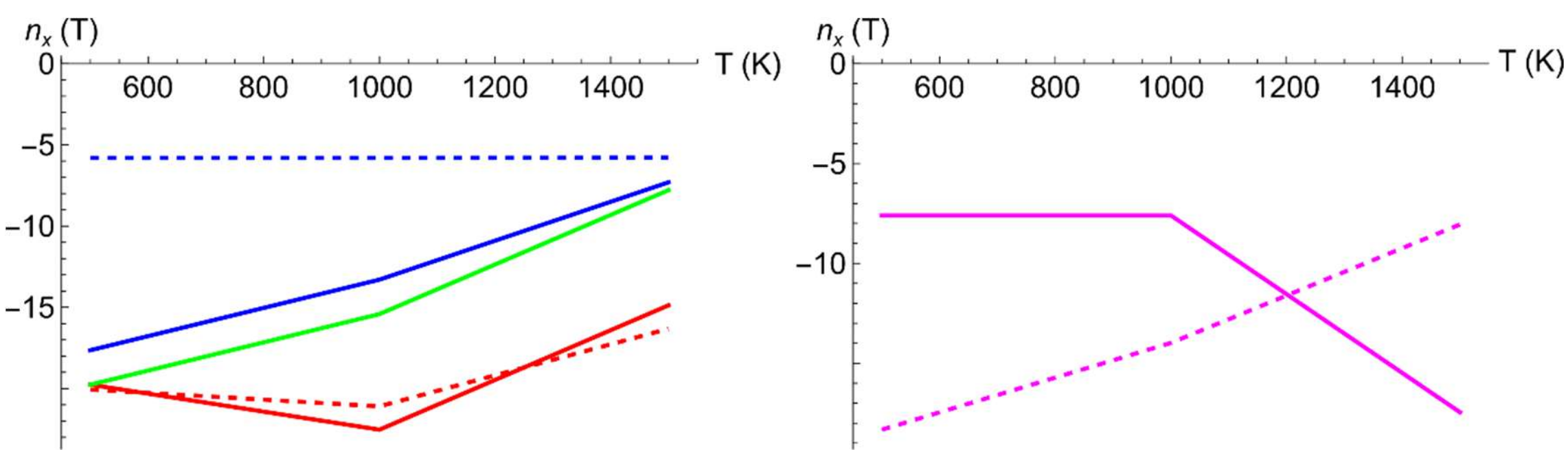

**Supplementary Figure 16**. Equilibrium distribution of species in the Kinetic Network. Shown is the base-10 logarithm of the molar fraction for selected chemical species at equilibrium in the kinetic network for 500, 1000 and 500 K. Left: H (blue), $H_2$ (dashed blue), C (red), $C_2$ (dashed red), Si (green). Right: $SiC_2$ (magenta, dashed), $SiH_4$ (magenta).

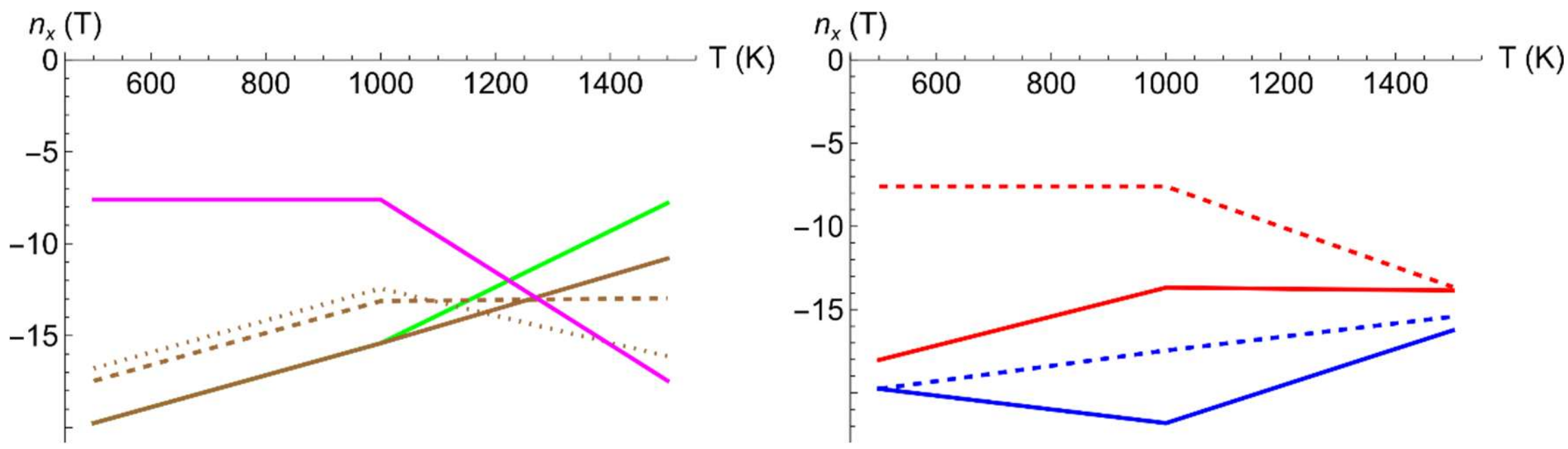

**Supplementary Figure 17**. Equilibrium distribution of species in the Kinetic Network. Shown is the base-10 logarithm of the molar fraction for selected chemical species at equilibrium in the kinetic network for 500, 1000 and 1500 K. Left: Si (green), SiH (brown), $SiH_2$ (brown, dashed), $SiH_3$ (brown, dotted), $SiH_4$ (magenta). Right: CH (blue), $CH_2$ (blue, dashed), $CH_3$ (red), $CH_4$ (red, dashed).

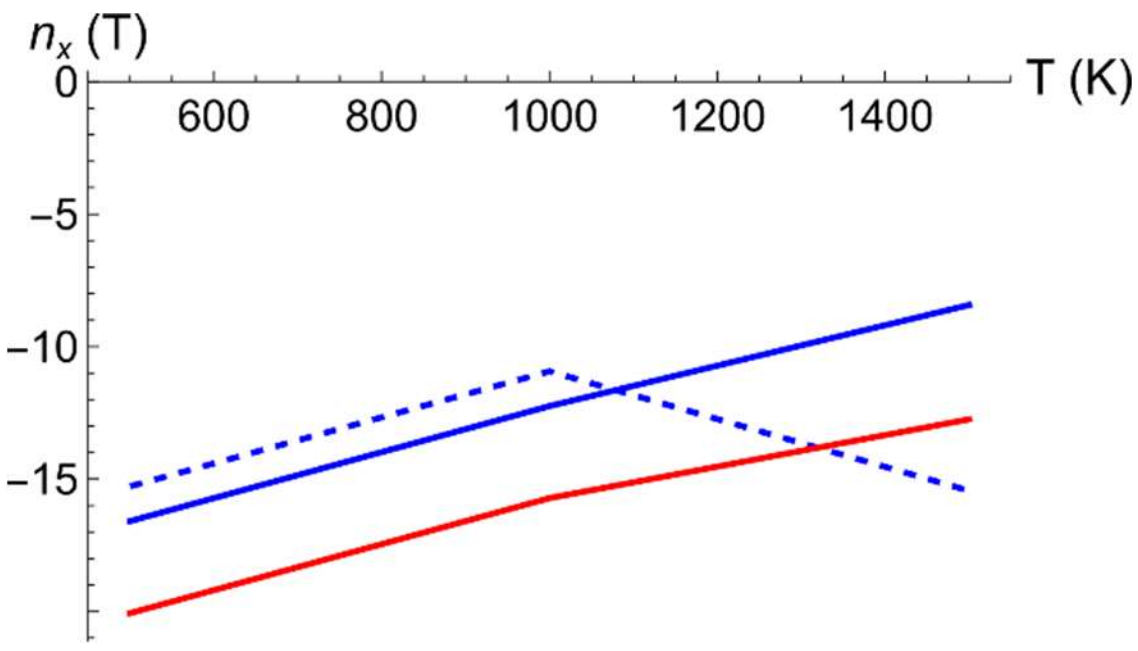

**Supplementary Figure 18**. Equilibrium distribution of species in the Kinetic Network. Shown is the base-10 logarithm of the molar fraction for selected chemical species at equilibrium in the kinetic network for 500, 1000 and 1500 K. $C_2H_2$ (blue), $C_2H_4$ (blue, dashed), $C_2H$ (red).

**Supplementary Section 12. From molecules to clusters and grains**

An important aspect in the context of silicon carbide dust formation is the growth process of dust grains from the molecular precursors and the plausible intermediate clusters linking molecules to dust. In particular, silicon carbide stoichiometric clusters $(SiC)_n$ have been suggested to play an important role as silicon carbide dust intermediates in C-rich stars [51] despite not being detected yet.

In the case of silicon carbide stoichiometric clusters, it has been shown that the most stable structures present atomic segregation whereas alternating Si-C bonds are not favoured [51, 52]. It has been proposed that these stoichiometric clusters form starting from a SiC molecule through the sequential addition of SiC molecules [51]. This growth mechanism will preserve the 1:1 stoichiometry in the dust grains. Early experiments by Frenklacht et al. found that silicon carbide small grains can efficiently grow at high temperatures and these small grains have been proposed as condensation sites for carbonaceous mantles [53, 54].

Yasuda & Kozasa [55] deepen in the nucleation process of silicon carbide by exploring both the Local Thermal Equilibrium (LTE) and non-LTE formation of silicon carbide grains. They assumed SiC as the starting molecule and considered a set of reactions with $Si_nC_m$ molecules leading to stoichiometric $(SiC)_n$ clusters. Additional reactions of $(SiC)_n$ with $C_2H_2$ and $C_2H$ are considered for the cluster growth but including Si atoms as reactants to preserve the 1:1 stoichiometry.

Our results suggest that a different scenario for silicon carbide dust growth might take place. First, we do not detect SiC but $SiC_2$ as the main molecule formed; thus, we identify $SiC_2$ as the most likely precursor of silicon carbide dust. This is also supported by astronomical observations [56]. Our kinetics modelling evidences that $SiC_2$ is formed from hydrocarbons and silane derivatives and the molecular content of our analogues as derived by LDI-MS experiments show that there is a large amount of hydrogenated clusters. In addition, the silicon carbide analogues are partially hydrogenated as revealed by XPS exhibiting Si-$CH_n$ moieties as shown by IRRAS. Altogether, our results indicate that nanosized dust might grow from $SiC_2$ through the interaction with hydrocarbons and silane derivatives and not only with $Si_nC_m$ molecules. Therefore, it would be interesting to explore chemical pathways that connect $SiC_2$ with hydrocarbons and silane derivatives for the growth of silicon carbide grains. For instance, we have calculated the Gibbs free energy of several possible chemical reactions (Supplementary Table 7). It can be observed that many of them are exergonic, particularly those involving $C_2H$, SiH and $SiH_4$. We note that $Si_2C_2H$ and $Si_2C_2H_2$ are detected in large amounts by LDI-MS in our analogues (see Supplementary Figure 10). Nevertheless, the investigation of plausible and complete reaction pathways leading to the growth of silicon carbide dust lies far beyond the scope of this manuscript.

**Supplementary Table 7**. Variation of the Gibbs-free energy for selected starting reactions invloving $SiC_2$ for the growth of silicon carbide dust

| Reaction | ΔG (kcal/mol) |
|---|---|
| $SiC_2 + SiC_2 \rightarrow Si_2C_2 + C_2$ | 60.42 |
| $SiC_2 + SiC_2 \rightarrow Si_2C_3 + C$ | 43.85 |
| $SiC_2 + C_2H \rightarrow SiC_4 + H$ | -28.58 |
| $SiC_2 + C_2H \rightarrow SiC_4H$ | -108.45 |
| $SiC_2 + C_2H_2 \rightarrow SiC_4 + H_2$ | -1.47 |
| $SiC_2 + C_2H_2 \rightarrow SiC_4H + H$ | 16.27 |
| $SiC_2 + SiH \rightarrow Si_2C_2 + H$ | -15.45 |
| $SiC_2 + SiH \rightarrow Si_2C_2H$ | -87.61 |
| $SiC_2 + SiH_4 \rightarrow Si_2C_2H_2 + H_2$ | -34.72 |

Finally, our results indicate that partially hydrogenated amorphous silicon carbide grains are primarily formed in the CSEs of C-rich AGBs. However, the presolar SiC grains found in primitive meteorites are crystalline, mainly in the β-polytype [57]. The transformation from hydrogenated amorphous grains into crystalline silicon carbide can occur through energetic processing. Both ion irradiation and thermal annealing has been shown to promote crystallization [58, 59] and dehydrogenation [60] of hydrogenated amorphous silicon carbide.

## Suplementary References